\documentclass[aps,amsfonts,pra,twocolumn,showpacs]{revtex4-1}
\usepackage{epsfig,amsmath,amssymb,bm,epsf,graphicx,psfrag}
\usepackage{color}

\newcommand{\bc}{\begin{center}}
\newcommand{\ec}{\end{center}}
\newcommand{\be}{\begin{equation}}
\newcommand{\ee}{\end{equation}}
\newcommand{\bea}{\begin{eqnarray}}
\newcommand{\eea}{\end{eqnarray}}

\newcommand{\mybox}[1]{		
	\begin{array}{|c|}
	\hline \!#1\!\\
	\hline
	\end{array}}

\newcommand{\ket}[1]{\left| #1\right\rangle}      
\newcommand{\bra}[1]{\left\langle #1\right|}      
\newcommand{\kets}[1]{| #1 \rangle}    				    

\newcommand{\bb}{ 
\setlength{\unitlength}{10bp}
\begin{picture}(0.3,1)
\put(0.15,-0.3){\line(0,1){1.2}}
\end{picture}
}

\newcommand{\blank}{ 
\kern-1bp
\setlength{\unitlength}{10bp}
\begin{picture}(1,1)
\put(0.15,0.05){$\bigcirc$}
\end{picture}
\kern+3bp
}

\newcommand{\qubit}{ 
\setlength{\unitlength}{10bp}
\kern+1bp
\begin{picture}(1,1)
\linethickness{1bp}
\put(0,-0.2){\line(0,1){1}}
\put(0,-0.2){\line(1,0){1}}
\put(1,0.8){\line(0,-1){1}}
\put(1,0.8){\line(-1,0){1}}
\end{picture}
\kern+1bp
}
\newcommand{\gate}{ 
\setlength{\unitlength}{10bp}
\kern+1bp
\begin{picture}(1,1)
\put(0.1,0){$\blacktriangleright$}
\linethickness{1bp}
\put(0,-0.2){\line(0,1){1}}
\put(0,-0.2){\line(1,0){1}}
\put(1,0.8){\line(0,-1){1}}
\put(1,0.8){\line(-1,0){1}}
\end{picture}
\kern+1bp
}

\newcommand{\plus}{\,+\,}   

\newcommand{\tur}{\circlearrowleft}   
\newcommand{\gat}{\,\blacktriangleright}  
\newcommand{\mov}{\,\vartriangleright}    

\newcommand{\movl}{\blacktriangleleft} 
\newcommand{\movle}{\vartriangleleft} 
\newcommand{\ff}{\rightarrow}
\newcommand{\ffp}{\Rightarrow}
\newcommand{\bul}{\,\bullet\,}       
\newcommand{\iga}{\:I\,}                  
\newcommand{\ici}{{\bigcirc \mkern-13mu   
	\mbox{\scriptsize{\textit{I}}}\mkern5mu\,}}
\newcommand{\wga}{W}						
\newcommand{\wci}{{\bigcirc \mkern-17mu 	
	\mbox{\scriptsize{\textit{W}}}\mkern1mu\,}}
\newcommand{\sga}{\:S\,}					
\newcommand{\sci}{{\bigcirc \mkern-14mu 	
	\mbox{\scriptsize{\textit{S}}}\mkern3mu\,}}
\newcommand{\aga}{\,A\,}					

\newcommand{\four}[4]{ 		
	\begin{array}{|c|c|}
	\hline #1 & #2 \\
	\hline #3 & #4 \\
	\hline
	\end{array}}
	
	\newcommand{\flour}[3]{ 		
	\begin{array}{|c|c|}
	\hline #1 & #2 \\
	\hline \multicolumn{2}{|c|}{#3} \\
	\hline
	\end{array}}
	
	\newcommand{\band}[2]{		
	\begin{array}{|r|r|}
	\hline #1 & #2 \\
	\hline
	\end{array}}
	
\newcommand{\TypeARule}[3]{
\begin{array}{ccc}
\!\!#1\!\! & & \!\!#2\!\!\\
\hline
\vline & \!\!#3\!\! &\vline\\
\hline
\end{array}
}

\newcommand{\TypeBRuleR}[3]{
\begin{array}{c}
 #1 \\
 \band{#2}{#3}
\end{array}
}

\newcommand{\TypeBRuleBB}[5]{
\begin{array}{c}#1\\
		\four{#2}{#3}{#4}{#5}
\end{array}
}

\newcommand{\TypeBRuleCC}[5]{
\begin{array}{c}#1\\
 \flour{#2}{#3}{#2(#4,#5)}
\end{array}
}

\newcommand{\TypeBRuleCD}[5]{
\begin{array}{c}#1\\
 \flour{#2}{#3}{#3^\dagger(#4,#5)}
\end{array}
}

\newcommand{\triURB}[4]{ 		
\begin{array}{c}
#1\\
	\begin{array}{r@{}c|c|}
	\cline{2-3} \vline & \:\: #2 & #3 \\ 
	\cline{2-3} & & #4 \\ 
	\cline{3-3} 
	\end{array}\:\:
	\end{array}}
	
	\newcommand{\triULB}[4]{ 		
	\begin{array}{c}
	#1\\
	\:\:\begin{array}{|c|c@{}l}
	\cline{1-2} #2 & #3 \:\: & \vline \\ 
	\cline{1-2} #4 & & \\ 
	\cline{1-1}
	\end{array}
	\end{array}}

\newlength{\onebox}
\begin{document}

\title{Hamiltonian quantum computer in one dimension}

\author{Tzu-Chieh Wei}
\affiliation{C. N. Yang Institute for Theoretical Physics and
Department of Physics and Astronomy, State University of New York at
Stony Brook, Stony Brook, NY 11794-3840, USA}

\author{John C. Liang}
\affiliation{Rumson-Fair Haven Regional High School, 74 Ridge Rd, Rumson, NJ 07760, USA}
\date{\today}

\begin{abstract}
Quantum computation can be achieved by preparing an appropriate initial product state of qudits and then letting it evolve under a fixed Hamiltonian. The readout is made by measurement on individual qudits at some later time. This approach is called the Hamiltonian quantum computation and  it includes, for example, the continuous-time quantum cellular automata and the universal quantum walk. We consider one spatial dimension and study the compromise between the locality $k$ and the local Hilbert space dimension $d$. For geometrically 2-local (i.e., $k=2$), it is known that $d=8$ is already sufficient for universal quantum computation but the Hamiltonian is not translationally invariant. As the locality $k$ increases, it is expected that the minimum required $d$ should decrease. We provide a construction of Hamiltonian quantum computer for $k=3$ with $d=5$.  One implication is that simulating 1D chains of spin-2 particles is BQP-complete. Imposing translation invariance will  increase the required $d$. For this we also construct another 3-local ($k=3$) Hamiltonian that is invariant under translation of a unit cell of two sites but that requires $d$ to be 8. 
\end{abstract}
\pacs{03.67.Lx, 03.67.-a, 
75.10.Jm}
\maketitle

\section{Introduction}
\begin{figure}
 \centering\includegraphics[width=0.48\textwidth]{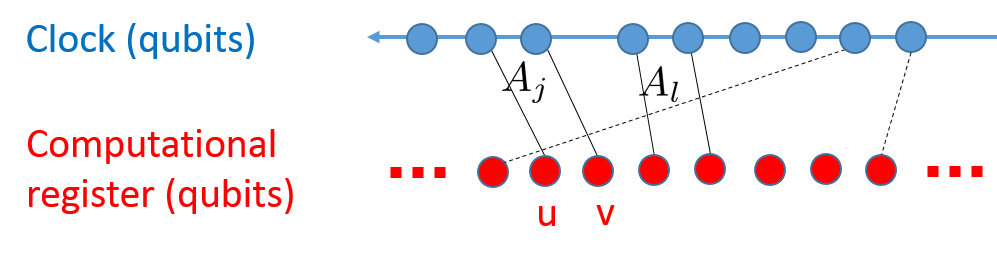}
        \caption{(color online) A schematic diagram for Feynman's Hamiltonian quantum computer. The top row of dots are qubits that constitute the clock. The bottom row of dots are qubits that constitute the computational part, i.e., those that one- and two-qubit gates are applied to. For example, the gate $A_j$ applies to qubits $u$ and $v$, indicated by two thin lines, when the two corresponding clock qubits have a change in their configuration $01\rightarrow 10$ (not shown). The interaction, given by the clock transition and gate operation, is highly non-local in general.
         \label{fig:HQC} }
\end{figure}

There are several approaches for quantum computation (QC), such as the standard circuit model~\cite{NielsenChuang}, topological QC~\cite{Kitaev,TQC}, adiabatic QC~\cite{Farhi}, measured-based QC~\cite{Oneway,Oneway2}, etc. In addition, quantum computation can be achieved by preparing an appropriate initial product state of qudits and then letting it evolve under a fixed Hamiltonian. The readout is made by measurement on individual qudits at some later time. Such an idea dated back to Benioff~\cite{Benioff} and Feynman~\cite{Feynman}. This is called a Hamiltonian quantum computer~\cite{NagajWocjan}. For example, Feynman provided an example Hamiltonian that is able to execute universal quantum computer, even though the interaction involves four particles residing on sites that are not geometrically local,
\begin{equation}
H_{\rm Feynman}= \sum_{j=0}^{k-1} \sigma_{j+1}^+\sigma_j^- A_{j+1} + {\rm h.c.}
\end{equation}
There are two important ingredients here; see Fig.~\ref{fig:HQC}. The first is the lowering and raising operators $\sigma^-$ and $\sigma^+$ that  act on a set of spin-1/2 particles, representing a discrete clock register. The clock state is initialized as $|00\dots001\rangle$ and, via the action of the Hamiltonian, can appear as $|00\dots 0 1_j 0\dots 0\rangle$ with one single excitation, giving rise to a unary representation of a discrete time. The second ingredient is the unitary gates $A_j$'s that represent all the gates (which can be one-qubit or two-qubit) that a quantum computer will apply to any qubits or qubit pairs that are in the computational register. If we denote the initial state of these qubits in the computational register as $|\psi_0\rangle$, then under the evolution of the Hamiltonian $e^{-i t H_{\rm Feynman}}$, the clock and computer register will be in a superposition of the following state
\begin{equation}
|\Psi(t)\rangle= \sum_j c_j(t)  |00\dots 0 1_j 0\dots0\rangle\otimes A_{j} A_{j-1}\dots A_1 |\psi_0\rangle,
\end{equation}
where the coefficients $c_j(t)$'s depend on the actual time $t$.  The time evolved state $|\Psi(t)\rangle$ thus contains states that represent any stage of quantum computation as gates are being applied to the initial state: $A_{j} A_{j-1}\dots A_1 |\psi_0\rangle$. The component of the computational register corresponding to the clock being $|10\dots 0\rangle$  gives the completion of the computation, i.e., all the gates have been applied: $A_{k} A_{k-1}\dots A_1 |\psi_0\rangle$. 
One can append many identity gates to the original gate sequence in order to boost the probability of ending up at a state where the desired quantum computation has been carried out. This gives  a general explanation why a Hamiltonian quantum computer can execute universal quantum computation. 

Feynman's idea was used and generalized by Kitaev to construct the so-called Local Hamiltonian Problems (LHP)~\cite{KitaevShenVyalyi}, which are concerned with the complexity of finding the ground-state energy. It turns out the 5-local LHP, which involves interacting terms of 5 particles that are not necessarily geometrically local, is believed to be a hard problem (called QMA in terms of complexity class) even for quantum computers~\cite{KitaevShenVyalyi}. The locality $k$ for QMA-complete LHP was, in a series of work, reduced to 2~\cite{KempeRegev,KempeKitaevRegev}, even with  nearest-neighbor interactions on two spatial dimensions~\cite{OliveiraTerhal}.
In one spatial dimension, it was shown by Aharonov et al. that 2-local 13-state Hamiltonians are QMA-complete~\cite{1DQMA}, and the local dimension $d$ is recently reduced to 8~\cite{Hallgren}. A key novelty in the one-dimensional case is that the use of qubits to represent the discrete clock was replaced by patterns of qudit configuration. This enables the reduction of interaction range to 2-local.

 In terms of one-dimensional Hamiltonian quantum computer, there have been various constructions, for example, the continuous-time quantum cellular automata by Vollbrecht and Cirac~\cite{VollbrechtCirac}, by Kay~\cite{Kay}, and by Nagaj and Wocjan~\cite{NagajWocjan}  as well as the universal quantum walk by Chase and Landahl~\cite{ChaseLandahl}. The 1D Hamiltonians in these constructions are nearest-neighbor two-body (or geometrically 2-local), but involve the dimension of local Hilbert space ranging from $d=8$~\cite{ChaseLandahl} and higher~\cite{VollbrechtCirac,Kay,NagajWocjan}. Here we study the compromise between the locality $k$ and the local dimension $d$ in one spatial dimension. As the locality $k$ increases, it is expected that the minimum required $d$ should decrease. For example, as a corollary of the results by Chase and Landahl, 6-local ($k=6$) qubit ($d=2$) Hamiltonians are universal, in the sense of quantum computation. For $k=4$, at most $d=3$ is needed for universality. But for $k=3$, how much lower than 8 can $d$ be? 
 
 The different constructions mentioned in previous paragraph share some common features: (i) the actual state of the computational register is represented by qubits  in a consecutive region of a larger array of qubits; (ii)  and there are parallel arrays of qudits that represent the program of the quantum computer, encoding instruction of gate movement and  operation~\cite{VollbrechtCirac,Kay,NagajWocjan,ChaseLandahl}. All of them give rise to translation invariant Hamiltonians, except Ref.~\cite{ChaseLandahl}. In this paper, we provide two constructions: (i) one that uses a 5-state 3-local Hamiltonian but is non-translation invariant, and (ii) 8-state 3-local Hamiltonian that is invariant under translation of a unit cell of two sites. The former is inspired by the design used in 1D QMA local Hamiltonian problems~\cite{1DQMA,Hallgren}, where the previous focus was on 2-locality whereas ours is on 3-locality instead. With the 2-locality relaxed to 3-locality, it is conceivable that the local Hilbert-space dimension can be reduced (e.g. from $d=9$ in Ref.~\cite{1DQMA}). Ours is an explicit demonstration of this. Our second construction is inspired by the translation invariant constructions in Refs.~\cite{VollbrechtCirac,Kay,NagajWocjan} and in particular the work by Nagaj and Wocjan~\cite{NagajWocjan}. We explicitly modify a particular scheme with $d=20$ in Ref.~\cite{NagajWocjan} and reduce $d$ to $8$. However, there was one scheme with $d=10$ in Ref.~\cite{NagajWocjan}, but we cannot reduce its $d$ further.  
  

The reason that our 5-state 3-local Hamiltonian is not translationally invariant (not under translation of finite lattice sites) is partly due to the site-dependence of gate operations (see Sec~\ref{sec:5state}), and hence it is not regarded as a quantum cellular automaton. 
One of the 2-local Hamiltonian automata of Nagaj and Wocjan uses the local dimension of $d=10$~\cite{NagajWocjan}. This implies the existence of a 4-local Hamiltonian automaton that requires $d=4$ but is invariant under translation of 2 sites, as well as a 6-local Hamiltonian automaton with $d=3$ that is invariant under translation of 3 sites. But this leaves open the question of 3-locality. Our previously mentioned $d=8$ construction thus gives an upper bound on the lowest $d$ in this case. However, it does not seem to be that much of an improvement, comparing to the 10-state 2-local construction of Nagaj and Wocjan~\cite{NagajWocjan}.

%

As summarized schematically in Fig.~\ref{fig:kvsd} and Fig.~\ref{fig:kvsdTI}, we consider one spatial dimension and focus on the continuous-time evolution of Hamiltonian and  focus on the compromise between the locality $k$ and the local dimension $d$.
But we mention that there were constructions using discrete-time quantum cellular automata, see e.g.~\cite{Lloyd,Watrous,Raussendorf,Shepherd} and as well as Hamiltonian quantum computer or quantum walk in two dimensions or higher, see e.g.~\cite{JanzingWocjan,Child}.
The remaining of the paper is organized as follows. In Sec.~\ref{sec:5state} we provide the 5-state 3-local construction and explain how the Hamiltonian is obtained. In Sec.~\ref{sec:8state} we consider translational invariance and construct 8-state 3-local transition rules that lead to a 8-state Hamiltonian invariant under translation of a unit cell of two sites. This construction can be regarded as a continuous-time quantum cellular automaton. In Sec.~\ref{sec:Prob} we analyze the probability of success. We conclude in Sec.~\ref{sec:con}.
\section{5-state non-translationally invariant construction}
\label{sec:5state}

\begin{figure}
 \includegraphics[width=0.5\textwidth]{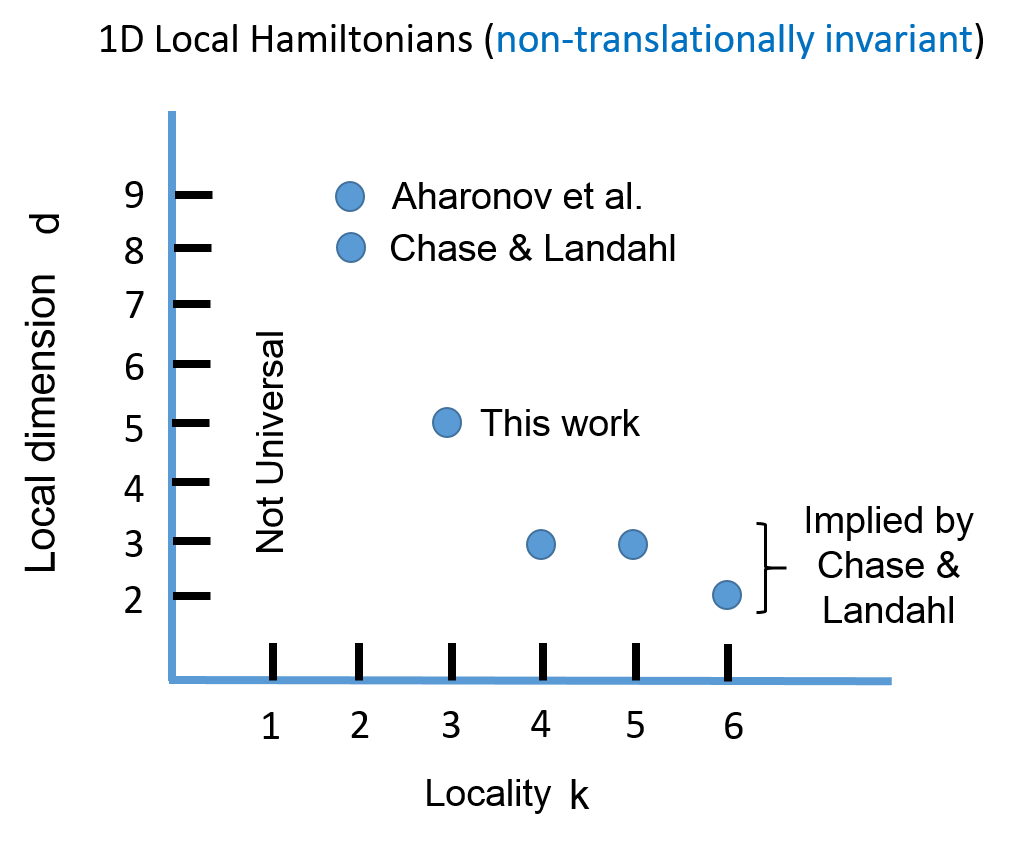}
        \caption{(color online) The status of locality $k$ vs. local Hilbert-space dimension (level) $d$ for universal quantum computation (BQP) in one spatial dimension.
         \label{fig:kvsd} }
\end{figure}
We are motivated to explore the compromise between the locality $k$ and the local dimension $d$ for 1D Hamiltonians capable of performing universal quantum computation. As explained in the Introduction, what remains to be answered is the 3-local case. Our construction here borrows the idea from 1D QMA local Hamiltonian problems~\cite{1DQMA,Hallgren} but is otherwise new,  and this adds a piece to the `phase diagram' illustrated in Fig.~\ref{fig:kvsd}. It turns out that a 2-local 9-state construction by Aharonov et al. for adiabatic QC can be used in the context of Hamiltonian QC~\cite{1DQMA}. The 8-state 1D QMA LHP by Hallgren et al.~\cite{Hallgren} actually uses effective transition rules involving 4 sites that are made from 2-local instructions. Our task here is not to find Hamiltonians that are QMA-complete, but to construct one that is universal for Hamiltonian quantum computer and that uses a small local dimension.
Our concern is 3-local whereas that of 1D QMA was 2-local~\cite{1DQMA}. Our transition rules are completely different.
  It turns out the local dimension we need is $d=5$ and we have two different types of sites.
The basis states on odd and even sites are, respectively,
\begin{eqnarray*}
\{\mov,\movle,\tur,\bul,\plus\}, \ \ \{\qubit^{[0]},\qubit^{[1]},\gate^{[0]},\gate^{[1]},\blank\}.
\end{eqnarray*}
(We can regard the system as consisting of the same kind of particles on all sites, but their interactions have two different preferred bases, depending on whether the site is even or odd.) There are two kinds of qubits: $\qubit$ and $\gate$, and the superscripts are used to indicate the logical qubit values (which will not be shown during the transitions below). The symbol $\blank$ is usually referred to as the unborn/dead symbol. The bullet $\bul$ and plus $\plus$ are used to space qubits as well as unborn/dead symbols. The empty left and right triangles indicate direction of movement. In addition, $\movle$ also serves to swap $\qubit$ and $\blank$. The turn-around symbol $\tur$ signals a change of direction. The complete transition rules are listed as follows and it is clear that the type of symbols implies the corresponding even or odd site.
\begin{eqnarray}\mbox{1:}\qquad
\begin{array}{lcr}
\gate\plus\qubit&\longrightarrow&U_m(\qubit\plus\gate)
\end{array}
\end{eqnarray}
\begin{eqnarray}\mbox{2:}\qquad
\begin{array}{lcr}
\gate\bul\bb\blank&\longrightarrow&\qubit\tur\bb\blank
\end{array}
\end{eqnarray}
\begin{eqnarray}\mbox{3:}\qquad
\begin{array}{lcr}
\tur\blank\bul&\longrightarrow&\movle\blank\bul
\end{array}
\end{eqnarray}
\begin{eqnarray}\mbox{4:}\qquad
\begin{array}{lcr}
\qubit\movle\blank&\longrightarrow&\blank\movle\qubit
\end{array}
\end{eqnarray}
\begin{eqnarray}\mbox{5a:}\qquad
\begin{array}{lcr}
\plus\blank\movle &\longrightarrow&\movle\blank\plus
\end{array}
\end{eqnarray}
\begin{eqnarray}\mbox{5b:}\qquad
\begin{array}{lcr}
\bul\blank\movle&\longrightarrow&\bul\blank\tur
\end{array}
\end{eqnarray}
\begin{eqnarray}\mbox{6a:}\qquad
\begin{array}{lcr}
\tur\bb\qubit\plus&\longrightarrow&\bul\bb\gate\plus
\end{array}
\end{eqnarray}
\begin{eqnarray}\mbox{6b:}\qquad
\begin{array}{lcr}
\tur:\!\qubit\plus&\longrightarrow&\bul\!:\!\qubit\mov
\end{array}
\end{eqnarray}
\begin{eqnarray}\mbox{7a:}\qquad
\begin{array}{lcr}
\mov\qubit\plus&\longrightarrow&\plus\qubit\mov
\end{array}
\end{eqnarray}
\begin{eqnarray}\mbox{7b:}\qquad
\begin{array}{lcr}
\mov\qubit\bul\!:&\longrightarrow&\plus\qubit\tur:
\end{array}
\end{eqnarray}

\begin{figure}
 \includegraphics[width=0.48\textwidth]{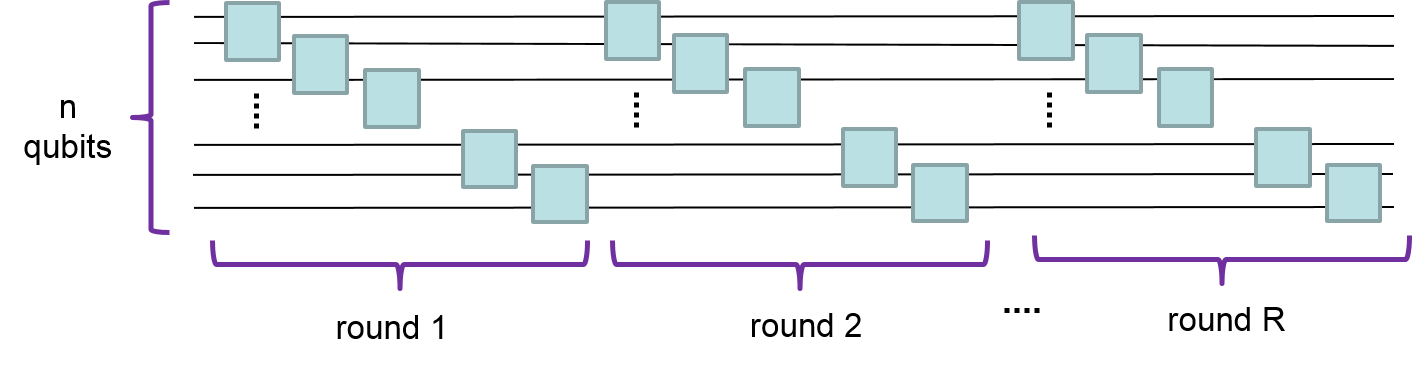}
        \caption{(color online) Universal circuit for quantum computation. This shows that gate sequence that will be simulated by our Hamiltonian quantum computers.
         \label{fig:circuit} }
\end{figure}

The reverse rules are obvious except that for Rule 1:
\begin{eqnarray}\mbox{1}^\dagger:\qquad
\begin{array}{lcr}
\qubit\plus\gate&\longrightarrow&U_m^\dagger(\gate\plus\qubit)
\end{array}
\end{eqnarray}
The gates $U_m$'s depend on the location and are exactly the gates used in a universal circuit model (see Fig.~\ref{fig:circuit}), where there are $R$ rounds of gate application, in each of which gates are applied sequentially between $i$-th and $(i+1)$-th qubits, for $i=1,\dots,n-1$. In terms of the order that they are applied, the gates are (from left to right)
\begin{equation}
	\underbrace{U_{1,1}, \dots, U_{1,n-1}}_{\textrm{1st round}},
	 \underbrace{U_{2,1}, \dots, U_{2,n-1}}_{\textrm{2nd round}},
	 \dots,
	\underbrace{U_{R,1}, \dots, U_{R,n-1}}_{\textrm{last round}}.
	\label{eqn:prog}
\end{equation} 
We note that the choice of the set of universal gates is arbitrary, for example, it can include the one-qubit gates such as the Hadamard gate and the $\pi/4$-gate, as well as the two-qubit  CNOT gate~\cite{NielsenChuang}, or even the $W$ and $S$ gates to be used in Sec.~\ref{sec:8state}. A one-qubit gate is a trivial special case of a two-qubit gate, where one of the two qubits is acted by an identity gate, and this is the reason that only two-qubit gates are drawn in Fig.~\ref{fig:circuit}.
Furthermore, our transition rules are inspired by those of Ref.~\cite{1DQMA} and use the same idea of data movement, i.e., since each gate is applied at a specific location, the particles need to be moved to the next round before the gate sequence for that round can take place. 

Let us use `$\big[$' and `$\big]$' to mark the geometric boundaries on the left and right sides, respectively. The symbol `$\bb$' marks the boundary between blocks and `$:$' marks the location to be {\it not\/} at the boundary. Except the first block which has only one site, all other blocks have $2n$ sites. There are $R+1$ blocks, where $R$ is the number of rounds for gate application.
Let us illustrate for the case of $n=3$ and $R=2$. The initial state is
\begin{eqnarray*}
[0]\qquad \big[\tur\bb\qubit\plus\qubit\plus\qubit\bul\bb\blank\bul\blank\bul\blank\bul\big]
\end{eqnarray*}
where the qubits $\qubit$'s are properly initialized. We will not show explicitly the logical values of the qubits and will refer to the list of symbols such as above as the {\it configuration\/}.
Then the rule 6a takes the configuration to
\begin{eqnarray*}
[1]\qquad \big[\bul\bb \gate\plus\qubit\plus\qubit\bul \bb \blank\bul\blank\bul\blank\bul\big]
\end{eqnarray*}
The gate $U_{1,1}$ will be applied and the configuration makes a transition via the rule 1 to
\begin{eqnarray*}
[2]\qquad \big[\bul\bb \qubit\plus\gate\plus\qubit\bul \bb \blank\bul\blank\bul\blank\bul\big]
\end{eqnarray*}
Similarly, the gate $U_{1,2}$ will be applied and the configuration makes a transition via the rule 1 to
\begin{eqnarray*}
[3]\qquad \big[\bul\bb \qubit\plus\qubit\plus\gate\bul \bb \blank\bul\blank\bul\blank\bul\big]
\end{eqnarray*}
Then via the rule 2, a turn-around symbol $\tur$ is generated:
\begin{eqnarray*}
[4]\qquad \big[\bul\bb \qubit\plus\qubit\plus\qubit\tur \bb \blank\bul\blank\bul\blank\bul\big]
\end{eqnarray*}
and followed by a generation of a left-moving symbol $\movle$ via the rule 3: 
\begin{eqnarray*}
[5]\qquad \big[\bul\bb \qubit\plus\qubit\plus\qubit\movle \bb \blank\bul\blank\bul\blank\bul\big]
\end{eqnarray*}
The qubit $\qubit$ and the unborn symbol $\blank$ swap via the rule 4:
\begin{eqnarray*}
[6]\qquad \big[\bul\bb \qubit\plus\qubit\plus\blank\movle \bb \qubit\bul\blank\bul\blank\bul\big]
\end{eqnarray*}
The left-moving symbol $\movle$ then swaps with the plus symbol $\plus$ via the rule 5a:
\begin{eqnarray*}
[7]\qquad \big[\bul\bb \qubit\plus\qubit\movle\blank\plus \bb \qubit\bul\blank\bul\blank\bul\big]
\end{eqnarray*}
The previous two steps repeat again:
\begin{eqnarray*}
[8]\qquad \big[\bul\bb \qubit\plus\blank\movle\qubit\plus \bb \qubit\bul\blank\bul\blank\bul\big]
\end{eqnarray*}
\begin{eqnarray*}
[9]\qquad \big[\bul\bb \qubit\movle\blank\plus\qubit\plus \bb \qubit\bul\blank\bul\blank\bul\big]
\end{eqnarray*}
The qubit $\qubit$ and the unborn symbol $\blank$ swaps via the rule 4:
\begin{eqnarray*}
[10]\qquad \big[\bul\bb \blank\movle\qubit\plus\qubit\plus \bb \qubit\bul\blank\bul\blank\bul\big]
\end{eqnarray*}
The left-moving symbol cannot move but generates a turn-around symbol $\tur$ via the rule 5b:
\begin{eqnarray*}
[11]\qquad \big[\bul\bb \blank\tur\qubit\plus\qubit\plus \bb \qubit\bul\blank\bul\blank\bul\big]
\end{eqnarray*}
This then creates a bullet $\bul$ and the right-moving symbol $\mov$ via the rule 6b:
\begin{eqnarray*}
[12]\qquad \big[\bul\bb \blank\bul\qubit\mov\qubit\plus \bb \qubit\bul\blank\bul\blank\bul\big]
\end{eqnarray*}
The right-moving symbol $\mov$ then swaps with the plus $\plus$ in front of it via the rule 7a:
\begin{eqnarray*}
[13]\qquad \big[\bul\bb \blank\bul\qubit\plus\qubit\mov \bb \qubit\bul\blank\bul\blank\bul\big]
\end{eqnarray*}
Then the right-moving symbol $\mov$ cannot move forward but causes a turn-around symbol $\tur$ to be generated via the rule 7b   (leaving behind a $\plus$ symbol):
\begin{eqnarray*}
[14]\qquad \big[\bul\bb \blank\bul\qubit\plus\qubit\plus \bb \qubit\tur\blank\bul\blank\bul\big]
\end{eqnarray*}
This then repeats and the process continues of moving the unborn symbol $\blank$ across the block of qubits to the left, as was shown previously in $[5]$ to $[13]$:
\begin{eqnarray*}
[15]\qquad \big[\bul\bb \blank\bul\qubit\plus\qubit\plus \bb \qubit\movle\blank\bul\blank\bul\big]
\end{eqnarray*}
\begin{eqnarray*}
[16]\qquad \big[\bul\bb \blank\bul\qubit\plus\qubit\plus \bb \blank\movle\qubit\bul\blank\bul\big]
\end{eqnarray*}
\begin{eqnarray*}
[17]\qquad \big[\bul\bb \blank\bul\qubit\plus\qubit\movle \bb \blank\plus\qubit\bul\blank\bul\big]
\end{eqnarray*}
\begin{eqnarray*}
[18]\qquad \big[\bul\bb \blank\bul\qubit\plus\blank\movle \bb \qubit\plus\qubit\bul\blank\bul\big]
\end{eqnarray*}
\begin{eqnarray*}
[19]\qquad \big[\bul\bb \blank\bul\qubit\movle\blank\plus \bb \qubit\plus\qubit\bul\blank\bul\big]
\end{eqnarray*}
\begin{eqnarray*}
[20]\qquad \big[\bul\bb \blank\bul\blank\movle\qubit\plus \bb \qubit\plus\qubit\bul\blank\bul\big]
\end{eqnarray*}
A turn-around symbol is then generated and there is a motion to the right:
\begin{eqnarray*}
[21]\qquad \big[\bul\bb \blank\bul\blank\tur\qubit\plus \bb \qubit\plus\qubit\bul\blank\bul\big]
\end{eqnarray*}
\begin{eqnarray*}
[22]\qquad \big[\bul\bb \blank\bul\blank\bul\qubit\mov \bb \qubit\plus\qubit\bul\blank\bul\big]
\end{eqnarray*}
\begin{eqnarray*}
[23]\qquad \big[\bul\bb \blank\bul\blank\bul\qubit\plus \bb \qubit\mov\qubit\bul\blank\bul\big]
\end{eqnarray*}
\begin{eqnarray*}
[24]\qquad \big[\bul\bb \blank\bul\blank\bul\qubit\plus \bb \qubit\plus\qubit\tur\blank\bul\big]
\end{eqnarray*}
After the turn-out symbol is generated, the whole process of transporting the unborn symbol $\blank$ repeats again:
\begin{eqnarray*}
[25]\qquad \big[\bul\bb \blank\bul\blank\bul\qubit\plus \bb \qubit\plus\qubit\movle\blank\bul\big]
\end{eqnarray*}
\begin{eqnarray*}
[26]\qquad \big[\bul\bb \blank\bul\blank\bul\qubit\plus \bb \qubit\plus\blank\movle\qubit\bul\big]
\end{eqnarray*}
\begin{eqnarray*}
[27]\qquad \big[\bul\bb \blank\bul\blank\bul\qubit\plus \bb \qubit\movle\blank\plus\qubit\bul\big]
\end{eqnarray*}
\begin{eqnarray*}
[28]\qquad \big[\bul\bb \blank\bul\blank\bul\qubit\plus \bb \blank\movle\qubit\plus\qubit\bul\big]
\end{eqnarray*}
\begin{eqnarray*}
[29]\qquad \big[\bul\bb \blank\bul\blank\bul\qubit\movle \bb \blank\plus\qubit\plus\qubit\bul\big]
\end{eqnarray*}
\begin{eqnarray*}
[30]\qquad \big[\bul\bb \blank\bul\blank\bul\blank\movle \bb \qubit\plus\qubit\plus\qubit\bul\big]
\end{eqnarray*}
Finally, we arrive at a similar state to the initial one, except that the whole qubit block has moved one block to the right and the gates $U_{2,1}$ and $U_{2,2}$ will be applied subsequently.
\begin{eqnarray*}
[31]\qquad \big[\bul\bb \blank\bul\blank\bul\blank\tur \bb \qubit\plus\qubit\plus\qubit\bul\big]
\end{eqnarray*}
\begin{eqnarray*}
[32]\qquad \big[\bul\bb \blank\bul\blank\bul\blank\bul \bb \gate\plus\qubit\plus\qubit\bul\big]
\end{eqnarray*}
\begin{eqnarray*}
[33]\qquad \big[\bul\bb \blank\bul\blank\bul\blank\bul \bb \qubit\plus\gate\plus\qubit\bul\big]
\end{eqnarray*}
After all the gates have been applied, the final state in the history of computation arrives:
\begin{eqnarray*}
[34]\qquad \big[\bul\bb \blank\bul\blank\bul\blank\bul \bb \qubit\plus\qubit\plus\gate\bul\big]
\end{eqnarray*}
The step number (counting from zero) is indicated in the square brackets $[\,\,\,]$ on the left in the above configurations. In general, for $n$ qubits with $R$ rounds, the total $T=(R-1)(3n^2+n)+n+1$.

What we have described above is the history of the computation via the transition rules (if the computation were run via discrete time). We will refer to the corresponding quantum states $|\psi_t\rangle$'s as the history states. However, the quantum computation will be executed not by discrete transitions, but via the continuous time evolution via the Hamiltonian: $\exp(-iHt)$. The procedure of running such  a Hamiltonian quantum computer is (1) prepare proper initial state, e.g. in $[0]$ above, (2) let it evolve under the Hamiltonian $H$, and (3) perform measurement in the computational basis at some time $\tau$. From the measurement outcome of all sites, we will be able to determine  at what stage of the computation the projected state is, and what the values of the qubits are. What remains to be shown is that the probability of arriving at a desired computation is high, which will be analyzed in Sec.~\ref{sec:Prob}. There, it will also be clear at what time the measurement should be taken (in fact, at a random moment).

\medskip\noindent {\bf Construction for the Hamiltonian}. We remark that these transition rules give rise to a Hamiltonian. For example, the rule 3 (including forward and backward) will contribute the following terms:
\begin{equation}
- \left|\movle\blank\bul\right\rangle \left\langle \tur\blank\bul\right|-\left|\tur\blank\bul\right\rangle\left\langle\movle\blank\bul\right|,
\end{equation}
 applicable at appropriate locations.
 For another example, the rule 6a will contribute the following terms,
 \begin{eqnarray}
- \sum_{s=0,1}& \left(\left|\bul\bb\gate^{[s]}\plus\right\rangle \left\langle \tur\bb\qubit^{[s]}\plus\right| \right.\nonumber\\
\,+\,&\left. \left|\tur\bb\qubit^{[s]}\plus\right\rangle\left\langle\bul\bb\gate^{[s]}\plus\right|\right),
\end{eqnarray}
where we have accounted for all possible qubit states.
For another example from the rule 1, we have the following terms 
 \begin{eqnarray}
&\!\!\!\!\!\!\!- \sum_{s} \left([{U_m}]^{{s_1'},{s_2'}}_{{s_1},{s_2}}\left|\qubit^{[s_1']}\plus\gate^{[s_2']}\right\rangle \left\langle \gate^{[s_1]}\plus\qubit^{[s_2]}\right| \right.\nonumber\\
&\!\!\!\!+\,\left. {[U_m^\dagger}]^{{s_1'},{s_2'}}_{{s_1},{s_2}} \left|\gate^{[s_1']}\plus\qubit^{[s_2']}\right\rangle\left\langle\qubit^{[s_1]}\plus\gate^{[s_2]}\right|\right).
\end{eqnarray}
As the construction for the Hamiltonian is clear, we will not explicitly write down all the terms. Moreover, it is the effective Hamiltonian in the basis of the history states that matters, which we discuss in Sec.~\ref{sec:Prob}.

\section{8-state translationally invariant construction}
\label{sec:8state}
\begin{figure}
 \includegraphics[width=0.5\textwidth]{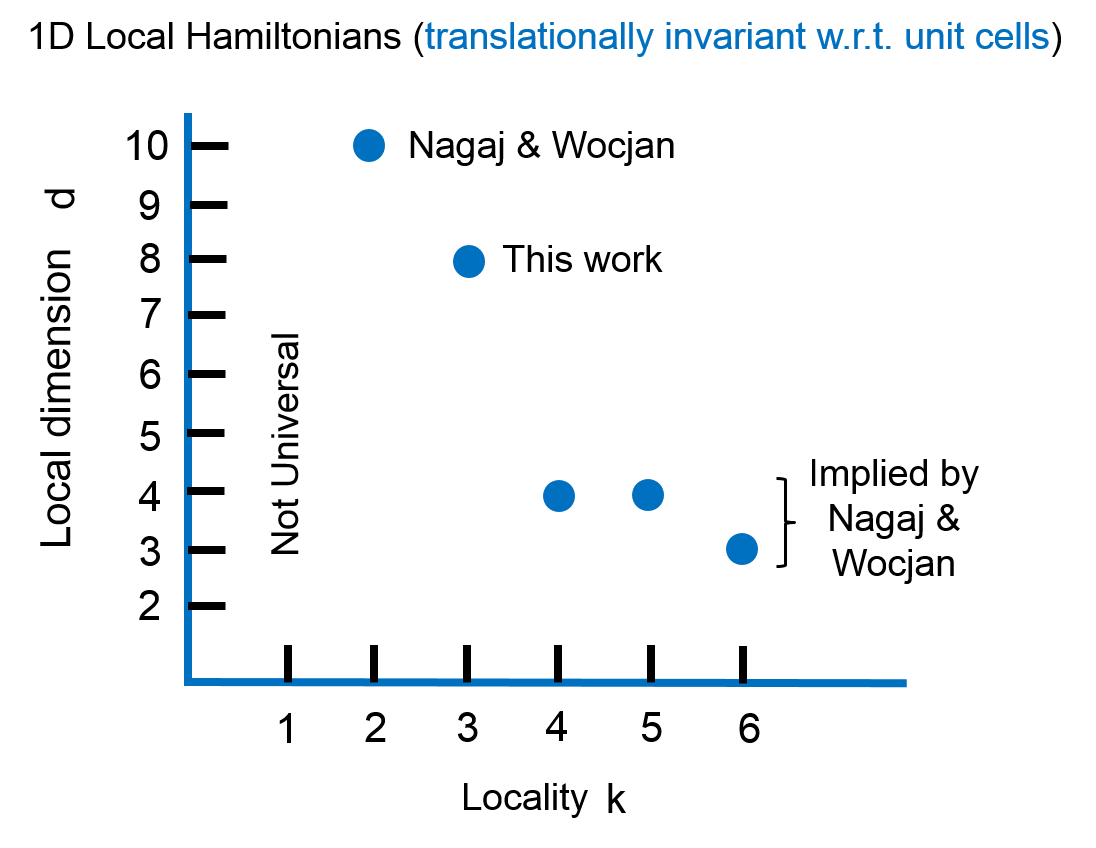}
        \caption{(color online) The status of locality $k$ vs. local Hilbert-space dimension (level) $d$ for universal quantum computation (BQP) in one spatial dimension. Here the Hamiltonian is restricted to translationally invariant ones.
         \label{fig:kvsdTI} }
\end{figure}

The 5-state construction in the last section yields a Hamiltonian that is not translationally invariant. If we impose the requirement that the resulting Hamiltonian to be translationally invariant (at least w.r.t. to translation across unit cells containing a fixed finite number of sites), then the minimum local dimension $d$ that is required to achieve a universal Hamiltonian quantum computer will be larger. In Fig.~\ref{fig:kvsdTI} we present the status of such 1D Hamiltonians on the $k$ vs $d$ diagram. It turns out that we can find a construction using a local dimension $d=8$. This can be regarded as the 3-local version of the continuous-time quantum cellular automata, and it has the advantage of serving as a programmable quantum computer. This quantum computer is also operated in three steps: (i) prepare appropriate initial product state, (ii) let the system evolve under the Hamiltonian, and (iii) perform measurement on all sites at a later time.

The transition rules used here are translationally invariant, as shown below. Our construction is a modification of the 2-local 20-state quantum cellular automaton by Nagaj and Wocjan~\cite{NagajWocjan}, and it also has a unique sequence of the history states via the transition rules on a properly initialized state. Similar to the construction in Sec.~\ref{sec:5state}, the dynamics of the transition rules (or the program) is such that
``particles'', mediating gate instructions, move above the data so that gates are executed at the right location. This data-movement technique comes
from the original construction in Ref.~\cite{1DQMA}.
On one sub-lattice (sites labeled by, say, half-integers), the local Hilbert space is composed of the following states
\begin{eqnarray}
\label{eqn:cursor}
\{\gat,\mov,\movl,\movle,\ffp,\ff,\tur,*\},
\end{eqnarray}
which can be regarded as the possible states of a cursor,
whereas on the other sub-lattice (sites labeled by, say, integers), it is composed of the following states
\begin{eqnarray}
\label{eqn:programdata}
\{\iga,\sga,\wga,\bul\}\otimes\{0,1\} ,
\end{eqnarray}
a tensor product of program and data registers.
Hence $d=8$ at every site. The one-dimensional physical lattice thus has two sites in a unit cell.
We note that the symbols $S$ and $W$ are associated with the swap gate $S$ and the $W$ gate, respectively, where
\begin{equation}
\label{eqn:S}
S=\left(
\begin{array}{cccc}
1 & 0 & 0 & 0 \\
0 & 0 & 1 & 0 \\
0 & 1 & 0 & 0 \\
0 & 0 & 0 & 1
\end{array}\right)
\end{equation}
and
\begin{equation}
\label{eqn:W}
W=\left(
\begin{array}{cccc}
1 & 0 & 0 & 0 \\
0 & 1 & 0 & 0 \\
0 & 0 & \frac{1}{\sqrt{2}} & \frac{-1}{\sqrt{2}} \\
0 & 0 &\frac{1}{\sqrt{2}} & \frac{1}{\sqrt{2}}
\end{array}\right),
\end{equation}
for which it is chosen for convenience that the control qubit is sitting geometrically to the left of the target qubit in our one-dimensional geometry.
One can show that $S$ and $W$ gates can simulate a universal set of gates (in fact only $W$ is needed, if it can be applied to any two qubits) and therefore $S$ and $W$ also constitute a set of universal gates; see Appendix~\ref{app:proof} for a proof.

In the original construction of Nagaj and Wocjan~\cite{NagajWocjan} on every site there are $d=20$ basis states given by
\begin{eqnarray*}
\{\iga,\sga,\wga,\bul,\ici,\sci,\wci,\gat,\mov,\tur\}\otimes\{0,1\}.
\end{eqnarray*}
To reduce the local dimension $d$ to $8$, we remove six of the symbols in the first bracket: $\{\ici,\sci,\wci,\gat,\mov,\tur\}$, leaving those in Eq.~(\ref{eqn:programdata}) composed of program and data registers. But to maintain the same computational capability we have to insert one additional site (referred to as the site of a cursor) with possible states shown in Eq.~(\ref{eqn:cursor}) in between every two original sites and modify the transition rules. It is based on their scheme that our scheme is constructed.

The transition rules of our Hamiltonian computer are listed as follows.
\begin{eqnarray}
\mbox{1a:} \qquad 
\begin{array}{lcr}
\TypeARule{\,*\,}{\movl}{\bul}&{}\atop\longrightarrow&\TypeARule{\movle}{\,*\,}{\bul}
\end{array}
\end{eqnarray}

\begin{eqnarray}
\mbox{1b:}\qquad
\begin{array}{lcr}
\TypeARule{\,*\,}{\movle}{\bul}&{}\atop\longrightarrow&\TypeARule{\tur}{\,*\,}{\bul}
\end{array}
\end{eqnarray}

\begin{eqnarray}
\mbox{1c:}\qquad
\begin{array}{lcr}
\TypeARule{\,*\,}{\movle}{\aga}&{}\atop\longrightarrow&\TypeARule{\movle}{\,*\,}{\aga}
\end{array}
\end{eqnarray}
We note that the gate symbol $\aga$ can be any one of the three: $\{\iga,\sga,\wga\}$, where $I$ is the idenity gate.
\begin{eqnarray}\mbox{2a:}\qquad
\begin{array}{lcr}
\TypeARule{\ffp}{\,*\,}{\bul}&{}\atop\longrightarrow&\TypeARule{\,*\,}{\gat}{\bul}
\end{array}
\end{eqnarray}

\begin{eqnarray}\mbox{2b:}\qquad
\begin{array}{lcr}
\TypeARule{\ff}{\,*\,}{\bul}&{}\atop\longrightarrow&\TypeARule{\,*\,}{\mov}{\bul}
\end{array}
\end{eqnarray}

\begin{eqnarray}\mbox{3a:}\qquad
\begin{array}{lcr}
\triURB{\tur}{\bul}{\bul}{\,1\,}&\longrightarrow&\triURB{\ffp}{\bul}{\bul}{1}
\end{array}
\end{eqnarray}

\begin{eqnarray}\mbox{3b:}\qquad
\begin{array}{lcr}
\triURB{\tur}{\bul}{\bul}{\,0\,}&\longrightarrow&\triURB{\ff}{\bul}{\bul}{0}
\end{array}
\end{eqnarray}

\begin{eqnarray}\mbox{4a:}\qquad
\begin{array}{lcr}
\TypeBRuleBB{\gat}{\bul}{\aga}{x}{y}&\longrightarrow&\TypeBRuleCC{\ffp}{\aga}{\bul}{x}{y}
\end{array}
\end{eqnarray}

\begin{eqnarray}\mbox{4b:}\qquad
\begin{array}{lcr}
\TypeBRuleR{\mov}{\bul}{\aga}&{}\atop\longrightarrow&\TypeBRuleR{\ff}{\aga}{\bul}
\end{array}
\end{eqnarray}

\begin{eqnarray}\mbox{5a:}\qquad
\begin{array}{lcr}
\triULB{\gat}{\bul}{\bul}{\,1\,}&\longrightarrow&\triULB{\movl}{\bul}{\bul}{1}
\end{array}
\end{eqnarray}

\begin{eqnarray}\mbox{5b:}\qquad
\begin{array}{lcr}
\triULB{\mov}{\bul}{\bul}{\,0\,}&\longrightarrow&\triULB{\movl}{\bul}{\bul}{0}
\end{array}
\end{eqnarray}

The reverse rules are obvious except that for Rule 4a:
\begin{eqnarray}\mbox{4a}^\dagger:\qquad
\begin{array}{lcr}
\TypeBRuleBB{\ffp}{\aga}{\bul}{x}{y}&\longrightarrow&\TypeBRuleCD{\gat}{\bul}{\aga}{x}{y}
\end{array}
\end{eqnarray}
\begin{widetext}

The total Hilbert space is composed of state of the following form:
\begin{eqnarray}
	\kets{\varphi} = \bigotimes_{j=1}^{L} \left(\ket{b_{j+\frac{1}{2}}}\otimes\ket{p_j} \otimes \ket{d_j}\right)_j,
	\label{start10product}
\end{eqnarray}
with $b_{j+\frac{1}{2}}$ denoting the state of the cursor register, $p_j$ the program register and $d_j$ the data register.

The initial state $|\psi_0\rangle$ is
\begin{eqnarray*}
[0]\qquad	\begin{array}{c|cccccccccccccccccccccccccccccccccc}	
		{j}  & 1 &\phantom{*} & 2 &\phantom{*} & 3 &\phantom{*}& 4 &\phantom{*}& 5 &\phantom{*}& 6&\phantom{*} &
		7 &\phantom{*}& 8 &\phantom{*}& 9 &\phantom{*}&  10 &\phantom{*}&
		11 &\phantom{*}& 12 &\phantom{*} & 13 &\phantom{*} & 14&\phantom{*}& 15 &\phantom{*}& \\
		\hline	
		b_{j+\frac{1}{2}}  & \phantom{\bul} &{*} &  \phantom{\bul} &{*} &  \phantom{\bul} &{*}&  \phantom{\bul} &{*}&  \phantom{\bul} &{*}&  \phantom{\bul}&{*} &
		 \phantom{\bul} &{*}& \phantom{\bul} &{*}&  \phantom{\bul} &{*}&
		 \phantom{\bul} &{*}&  \phantom{\bul} &{*} &  \phantom{\bul} &{*} & \phantom{\bul}&{*}& \phantom{\bul} &{*} &  \phantom{\bul} &{\blacktriangleleft} \\
		\hline
		p_{j}  & \bul &\phantom{*} & \bul &\phantom{*} & \bul &\phantom{*}& \bul &\phantom{*}& \bul &\phantom{*}& \bul&\phantom{*} &
		\iga &\phantom{*}& \wga &\phantom{*}& \sga &\phantom{*}&  \iga &\phantom{*}&
		\iga &\phantom{*}& \sga &\phantom{*} & \wga &\phantom{*} & \iga&\phantom{*}& \bul &\phantom{*}&
		\\		
		d_{j}  & 0  &\phantom{*}    & 1  &\phantom{*}     & 0   &\phantom{*}  & 0   &\phantom{*}  & 0   &\phantom{*} &1   &\phantom{*}  & 
		w_1  &\phantom{*} & w_2  &\phantom{*} & w_3  &\phantom{*} & 1   &\phantom{*}&
		0   &\phantom{*}  & 0   &\phantom{*}  & 0 &\phantom{*}   & 1   &\phantom{*}    	& 0   &\phantom{*}	
 	\end{array} 
\end{eqnarray*}
where $w_i$'s correspond to the actual qubits in the circuit model (see Fig.~\ref{fig:circuit}) and needs to be properly initialized. We also note that the gates in the program register $p_j$'s  are arranged in the order
\begin{eqnarray}
	I , \underbrace{U_{1,1}, \dots, U_{1,n-1}}_{\textrm{1st round}},
	I, I, \underbrace{U_{2,1}, \dots, U_{2,n-1}}_{\textrm{2nd round}},
	I, \dots,
	I, \underbrace{U_{R,1}, \dots U_{R,n-1}}_{\textrm{last round}},\,I ,
	\label{prog}
\end{eqnarray} 
with $U_{k,j}$ being one of the three possible gates in the set $\{W,S,I\}$,  and, in comparison with the gate sequence in Eq.~(\ref{eqn:prog}), here each round of gates is both preceded and followed by an identity gate.  It is important to add the extra identity gates $I$'s so that  there will not be any undesired gate operation between the qubit $\omega_1$ and the qubit to its left nor between the qubit $\omega_n$ and the qubit to its right~\cite{NagajWocjan}. Moreover, the qubit pattern in the data register, as illustrated in the example above, is
\begin{equation}
 0\underbrace{100\dots0}_{ (R\!-\!1)\, \textrm{blocks}} 1\,\omega_1\omega_2\omega_3\dots\omega_n\underbrace{100\dots0}_{(R\!-\!1)\, \textrm{ blocks}}10,
 \end{equation} where the spacing between the $1$'s being the same as the number of gates (including the identity gates) in a round. We note that there needs to be at least $(R-1)$ blocks of space to the left of all gates, so that $R$ rounds of gates can be completely executed.  The particular pattern was designed by Nagaj and Wocjan~\cite{NagajWocjan} to appropriately activate gate operations. Via the rule 1a, the initial state makes a transition to the following
\begin{eqnarray*}
[1] \qquad	\begin{array}{c|cccccccccccccccccccccccccccccccccc}	
		b_{j+\frac{1}{2}}  & \phantom{\bul} &{*} &  \phantom{\bul} &{*} &  \phantom{\bul} &{*}&  \phantom{\bul} &{*}&  \phantom{\bul} &{*}&  \phantom{\bul}&{*} &
		 \phantom{\bul} &{*}& \phantom{\bul} &{*}&  \phantom{\bul} &{*}&
		 \phantom{\bul} &{*}&  \phantom{\bul} &{*} &  \phantom{\bul} &{*} & \phantom{\bul}&{*}& \phantom{\bul} &{\movle}&  \phantom{\bul} & {*} \\
		\hline
		p_{j}  & \bul &\phantom{*} & \bul &\phantom{*} & \bul &\phantom{*}& \bul &\phantom{*}& \bul &\phantom{*}& \bul&\phantom{*} &
		\iga &\phantom{*}& \wga &\phantom{*}& \sga &\phantom{*}&  \iga &\phantom{*}&
		\iga &\phantom{*}& \sga &\phantom{*} & \wga &\phantom{*} & \iga&\phantom{*}& \bul &\phantom{*}&
		\\		
		d_{j}  & 0  &\phantom{*}    & 1  &\phantom{*}     & 0   &\phantom{*}  & 0   &\phantom{*}  & 0   &\phantom{*} &1   &\phantom{*}  & 
		w_1  &\phantom{*} & w_2  &\phantom{*} & w_3  &\phantom{*} & 1   &\phantom{*}&
		0   &\phantom{*}  & 0   &\phantom{*}  & 0 &\phantom{*}   & 1   &\phantom{*}    	& 0   &\phantom{*}	
 	\end{array} 
\end{eqnarray*}
where the solid left-triangle moves one step forward and turns into a empty left-triangle.  Via the rule 1c, the empty left-tirangle moves one step forward: 
\begin{eqnarray*}
[2] \qquad	\begin{array}{c|cccccccccccccccccccccccccccccccccc}	
		b_{j+\frac{1}{2}}  & \phantom{\bul} &{*} &  \phantom{\bul} &{*} &  \phantom{\bul} &{*}&  \phantom{\bul} &{*}&  \phantom{\bul} &{*}&  \phantom{\bul}&{*} &
		 \phantom{\bul} &{*}& \phantom{\bul} &{*}&  \phantom{\bul} &{*}&
		 \phantom{\bul} &{*}&  \phantom{\bul} &{*} &  \phantom{\bul} &{*} & \phantom{\bul}&{\movle}& \phantom{\bul} &{*} &  \phantom{\bul} & {*} \\
		\hline
		p_{j}  & \bul &\phantom{*} & \bul &\phantom{*} & \bul &\phantom{*}& \bul &\phantom{*}& \bul &\phantom{*}& \bul&\phantom{*} &
		\iga &\phantom{*}& \wga &\phantom{*}& \sga &\phantom{*}&  \iga &\phantom{*}&
		\iga &\phantom{*}& \sga &\phantom{*} & \wga &\phantom{*} & \iga&\phantom{*}& \bul &\phantom{*}&
		\\		
		d_{j}  & 0  &\phantom{*}    & 1  &\phantom{*}     & 0   &\phantom{*}  & 0   &\phantom{*}  & 0   &\phantom{*} &1   &\phantom{*}  & 
		w_1  &\phantom{*} & w_2  &\phantom{*} & w_3  &\phantom{*} & 1   &\phantom{*}&
		0   &\phantom{*}  & 0   &\phantom{*}  & 0 &\phantom{*}   & 1   &\phantom{*}    	& 0   &\phantom{*}	
 	\end{array} 
\end{eqnarray*}
and continues until the configuration becomes the following one:
\begin{eqnarray*}
[9]\qquad	\begin{array}{c|cccccccccccccccccccccccccccccccccc}	
		b_{j+\frac{1}{2}}  & \phantom{\bul} &{*} &  \phantom{\bul} &{*} &  \phantom{\bul} &{*}&  \phantom{\bul} &{*}&  \phantom{\bul} &{*}&  \phantom{\bul}&{\movle} &
		 \phantom{\bul} &{*}& \phantom{\bul} &{*}&  \phantom{\bul} &{*}&
		 \phantom{\bul} &{*}&  \phantom{\bul} &{*} &  \phantom{\bul} &{*} & \phantom{\bul}&{*}& \phantom{\bul} &{*} &  \phantom{\bul} & {*} \\
		\hline
		p_{j}  & \bul &\phantom{*} & \bul &\phantom{*} & \bul &\phantom{*}& \bul &\phantom{*}& \bul &\phantom{*}& \bul&\phantom{*} &
		\iga &\phantom{*}& \wga &\phantom{*}& \sga &\phantom{*}&  \iga &\phantom{*}&
		\iga &\phantom{*}& \sga &\phantom{*} & \wga &\phantom{*} & \iga&\phantom{*}& \bul &\phantom{*}&
		\\		
		d_{j}  & 0  &\phantom{*}    & 1  &\phantom{*}     & 0   &\phantom{*}  & 0   &\phantom{*}  & 0   &\phantom{*} &1   &\phantom{*}  & 
		w_1  &\phantom{*} & w_2  &\phantom{*} & w_3  &\phantom{*} & 1   &\phantom{*}&
		0   &\phantom{*}  & 0   &\phantom{*}  & 0 &\phantom{*}   & 1   &\phantom{*}    	& 0   &\phantom{*}	
 	\end{array} 
\end{eqnarray*}
The rule 1b then kicks in, generating a turn-around symbol $\tur$:
\begin{eqnarray*}
	[10]\qquad \begin{array}{c|cccccccccccccccccccccccccccccccccc}	
		b_{j+\frac{1}{2}}  & \phantom{\bul} &{*} &  \phantom{\bul} &{*} &  \phantom{\bul} &{*}&  \phantom{\bul} &{*}&  \phantom{\bul} &{\tur}&  \phantom{\bul}&{*} &
		 \phantom{\bul} &{*}& \phantom{\bul} &{*}&  \phantom{\bul} &{*}&
		 \phantom{\bul} &{*}&  \phantom{\bul} &{*} &  \phantom{\bul} &{*} & \phantom{\bul}&{*}& \phantom{\bul} &{*} &  \phantom{\bul} & {*} \\
		\hline
		p_{j}  & \bul &\phantom{*} & \bul &\phantom{*} & \bul &\phantom{*}& \bul &\phantom{*}& \bul &\phantom{*}& \bul&\phantom{*} &
		\iga &\phantom{*}& \wga &\phantom{*}& \sga &\phantom{*}&  \iga &\phantom{*}&
		\iga &\phantom{*}& \sga &\phantom{*} & \wga &\phantom{*} & \iga&\phantom{*}& \bul &\phantom{*}&
		\\		
		d_{j}  & 0  &\phantom{*}    & 1  &\phantom{*}     & 0   &\phantom{*}  & 0   &\phantom{*}  & 0   &\phantom{*} &1   &\phantom{*}  & 
		w_1  &\phantom{*} & w_2  &\phantom{*} & w_3  &\phantom{*} & 1   &\phantom{*}&
		0   &\phantom{*}  & 0   &\phantom{*}  & 0 &\phantom{*}   & 1   &\phantom{*}    	& 0   &\phantom{*}	
 	\end{array} 
\end{eqnarray*}
Via the rule 3a, the turn-around symbol becomes a double right-arrow:
\begin{eqnarray*}
[11]\qquad	\begin{array}{c|cccccccccccccccccccccccccccccccccc}	
		b_{j+\frac{1}{2}}  & \phantom{\bul} &{*} &  \phantom{\bul} &{*} &  \phantom{\bul} &{*}&  \phantom{\bul} &{*}&  \phantom{\bul} &{\ffp}&  \phantom{\bul}&{*} &
		 \phantom{\bul} &{*}& \phantom{\bul} &{*}&  \phantom{\bul} &{*}&
		 \phantom{\bul} &{*}&  \phantom{\bul} &{*} &  \phantom{\bul} &{*} & \phantom{\bul}&{*}& \phantom{\bul} &{*} &  \phantom{\bul} & {*} \\
		\hline
		p_{j}  & \bul &\phantom{*} & \bul &\phantom{*} & \bul &\phantom{*}& \bul &\phantom{*}& \bul &\phantom{*}& \bul&\phantom{*} &
		\iga &\phantom{*}& \wga &\phantom{*}& \sga &\phantom{*}&  \iga &\phantom{*}&
		\iga &\phantom{*}& \sga &\phantom{*} & \wga &\phantom{*} & \iga&\phantom{*}& \bul &\phantom{*}&
		\\		
		d_{j}  & 0  &\phantom{*}    & 1  &\phantom{*}     & 0   &\phantom{*}  & 0   &\phantom{*}  & 0   &\phantom{*} &1   &\phantom{*}  & 
		w_1  &\phantom{*} & w_2  &\phantom{*} & w_3  &\phantom{*} & 1   &\phantom{*}&
		0   &\phantom{*}  & 0   &\phantom{*}  & 0 &\phantom{*}   & 1   &\phantom{*}    	& 0   &\phantom{*}	
 	\end{array} 
\end{eqnarray*}
The double right-arrow moves and makes a transition via the rule 2a into a solid right-triangle:  
\begin{eqnarray*}
[12]\qquad	\begin{array}{c|cccccccccccccccccccccccccccccccccc}	
		b_{j+\frac{1}{2}}  & \phantom{\bul} &{*} &  \phantom{\bul} &{*} &  \phantom{\bul} &{*}&  \phantom{\bul} &{*}&  \phantom{\bul} &{*}&  \phantom{\bul}&{\gat} &
		 \phantom{\bul} &{*}& \phantom{\bul} &{*}&  \phantom{\bul} &{*}&
		 \phantom{\bul} &{*}&  \phantom{\bul} &{*} &  \phantom{\bul} &{*} & \phantom{\bul}&{*}& \phantom{\bul} &{*} &  \phantom{\bul} & {*} \\
		\hline
		p_{j}  & \bul &\phantom{*} & \bul &\phantom{*} & \bul &\phantom{*}& \bul &\phantom{*}& \bul &\phantom{*}& \bul&\phantom{*} &
		\iga &\phantom{*}& \wga &\phantom{*}& \sga &\phantom{*}&  \iga &\phantom{*}&
		\iga &\phantom{*}& \sga &\phantom{*} & \wga &\phantom{*} & \iga&\phantom{*}& \bul &\phantom{*}&
		\\		
		d_{j}  & 0  &\phantom{*}    & 1  &\phantom{*}     & 0   &\phantom{*}  & 0   &\phantom{*}  & 0  &\phantom{*} &\mybox{1}   &\phantom{*}  & 
		\mybox{w_1}  &\phantom{*} & w_2  &\phantom{*} & w_3  &\phantom{*} & 1   &\phantom{*}&
		0   &\phantom{*}  & 0   &\phantom{*}  & 0 &\phantom{*}   & 1   &\phantom{*}    	& 0   &\phantom{*}	
 	\end{array} 
\end{eqnarray*}
This is where the gate $\iga$ is applied and a double right-arrow is generated:
\begin{eqnarray*}
[13]\qquad	\begin{array}{c|cccccccccccccccccccccccccccccccccc}	
		b_{j+\frac{1}{2}}  & \phantom{\bul} &{*} &  \phantom{\bul} &{*} &  \phantom{\bul} &{*}&  \phantom{\bul} &{*}&  \phantom{\bul} &{*}&  \phantom{\bul}&{\ffp} &
		 \phantom{\bul} &{*}& \phantom{\bul} &{*}&  \phantom{\bul} &{*}&
		 \phantom{\bul} &{*}&  \phantom{\bul} &{*} &  \phantom{\bul} &{*} & \phantom{\bul}&{*}& \phantom{\bul} &{*} &  \phantom{\bul} & {*} \\
		\hline
		p_{j}  & \bul &\phantom{*} & \bul &\phantom{*} & \bul &\phantom{*}& \bul &\phantom{*}& \bul &\phantom{*}& \iga&\phantom{*} &
		\bul &\phantom{*}& \wga &\phantom{*}& \sga &\phantom{*}&  \iga &\phantom{*}&
		\iga &\phantom{*}& \sga &\phantom{*} & \wga &\phantom{*} & \iga&\phantom{*}& \bul &\phantom{*}&
		\\		
		d_{j}  & 0  &\phantom{*}    & 1  &\phantom{*}     & 0   &\phantom{*}  & 0   &\phantom{*}  & 0   &\phantom{*} &1   &\phantom{*}  & 
		w_1  &\phantom{*} & w_2  &\phantom{*} & w_3  &\phantom{*} & 1   &\phantom{*}&
		0   &\phantom{*}  & 0   &\phantom{*}  & 0 &\phantom{*}   & 1   &\phantom{*}    	& 0   &\phantom{*}	
 	\end{array} 
\end{eqnarray*} 
Note that for convenience, we will not change the symbols $\omega_i$'s even if nontrivial gate effect arises. The two boxes indicate where the two qubits were affected by the gate operation.
The previous two steps repeat a few times, but note that the gates will have no effect outside the qubit block $\omega_i$'s, and we arrive at the following configuration:
\begin{eqnarray*}
[26]\qquad	\begin{array}{c|cccccccccccccccccccccccccccccccccc}	
		b_{j+\frac{1}{2}}  & \phantom{\bul} &{*} &  \phantom{\bul} &{*} &  \phantom{\bul} &{*}&  \phantom{\bul} &{*}&  \phantom{\bul} &{*}&  \phantom{\bul}&{*} &
		 \phantom{\bul} &{*}& \phantom{\bul} &{*}&  \phantom{\bul} &{*}&
		 \phantom{\bul} &{*}&  \phantom{\bul} &{*} &  \phantom{\bul} &{*} & \phantom{\bul}&{\gat}& \phantom{\bul} &{*} &  \phantom{\bul} & {*} \\
		\hline
		p_{j}  & \bul &\phantom{*} & \bul &\phantom{*} & \bul &\phantom{*}& \bul &\phantom{*}& \bul &\phantom{*}& \iga&\phantom{*} & \wga &\phantom{*}& \sga &\phantom{*}&  \iga &\phantom{*}&
		\iga &\phantom{*}& \sga &\phantom{*} & \wga &\phantom{*} & 
		\bul &\phantom{*}&\iga&\phantom{*}& \bul &\phantom{*}&
		\\		
		d_{j}  & 0  &\phantom{*}    & 1  &\phantom{*}     & 0   &\phantom{*}  & 0   &\phantom{*}  & 0   &\phantom{*} &1   &\phantom{*}  & 
		w_1  &\phantom{*} & w_2  &\phantom{*} & w_3  &\phantom{*} & 1   &\phantom{*}&
		0   &\phantom{*}  & 0   &\phantom{*}  & \mybox{0} &\phantom{*}   & \mybox{1}   &\phantom{*}    	& 0   &\phantom{*}	
 	\end{array} 
\end{eqnarray*}
It then transits (via the rule 4a) into
\begin{eqnarray*}
[27]\qquad	\begin{array}{c|cccccccccccccccccccccccccccccccccc}	
		b_{j+\frac{1}{2}}  & \phantom{\bul} &{*} &  \phantom{\bul} &{*} &  \phantom{\bul} &{*}&  \phantom{\bul} &{*}&  \phantom{\bul} &{*}&  \phantom{\bul}&{*} &
		 \phantom{\bul} &{*}& \phantom{\bul} &{*}&  \phantom{\bul} &{*}&
		 \phantom{\bul} &{*}&  \phantom{\bul} &{*} &  \phantom{\bul} &{*} & \phantom{\bul}&{\ffp}& \phantom{\bul} &{*} &  \phantom{\bul} & {*} \\
		\hline
		p_{j}  & \bul &\phantom{*} & \bul &\phantom{*} & \bul &\phantom{*}& \bul &\phantom{*}& \bul &\phantom{*}& \iga&\phantom{*} & \wga &\phantom{*}& \sga &\phantom{*}&  \iga &\phantom{*}&
		\iga &\phantom{*}& \sga &\phantom{*} & \wga &\phantom{*} & 
		\iga &\phantom{*}&\bul&\phantom{*}& \bul &\phantom{*}&
		\\		
		d_{j}  & 0  &\phantom{*}    & 1  &\phantom{*}     & 0   &\phantom{*}  & 0   &\phantom{*}  & 0   &\phantom{*} &1   &\phantom{*}  & 
		w_1  &\phantom{*} & w_2  &\phantom{*} & w_3  &\phantom{*} & 1   &\phantom{*}&
		0   &\phantom{*}  & 0   &\phantom{*}  & 0 &\phantom{*}   & 1   &\phantom{*}    	& 0   &\phantom{*}	
 	\end{array} 
\end{eqnarray*}
The double right-arrow moves and becomes the solid right-triangle (via the rule 2a):
\begin{eqnarray*}
[28]\qquad	\begin{array}{c|cccccccccccccccccccccccccccccccccc}	
		b_{j+\frac{1}{2}}  & \phantom{\bul} &{*} &  \phantom{\bul} &{*} &  \phantom{\bul} &{*}&  \phantom{\bul} &{*}&  \phantom{\bul} &{*}&  \phantom{\bul}&{*} &
		 \phantom{\bul} &{*}& \phantom{\bul} &{*}&  \phantom{\bul} &{*}&
		 \phantom{\bul} &{*}&  \phantom{\bul} &{*} &  \phantom{\bul} &{*} & \phantom{\bul}&{*}& \phantom{\bul} &{\gat} &  \phantom{\bul} & {*} \\
		\hline
		p_{j}  & \bul &\phantom{*} & \bul &\phantom{*} & \bul &\phantom{*}& \bul &\phantom{*}& \bul &\phantom{*}& \iga&\phantom{*} & \wga &\phantom{*}& \sga &\phantom{*}&  \iga &\phantom{*}&
		\iga &\phantom{*}& \sga &\phantom{*} & \wga &\phantom{*} & 
		\iga &\phantom{*}&\bul&\phantom{*}& \bul &\phantom{*}&
		\\		
		d_{j}  & 0  &\phantom{*}    & 1  &\phantom{*}     & 0   &\phantom{*}  & 0   &\phantom{*}  & 0   &\phantom{*} &1   &\phantom{*}  & 
		w_1  &\phantom{*} & w_2  &\phantom{*} & w_3  &\phantom{*} & 1   &\phantom{*}&
		0   &\phantom{*}  & 0   &\phantom{*}  & 0 &\phantom{*}   & 1   &\phantom{*}    	& 0   &\phantom{*}	
 	\end{array} 
\end{eqnarray*}
At this moment the solid right-triangle makes a transition (via the rule 5a) to a solid left-triangle:
\begin{eqnarray*}
[29]\qquad	\begin{array}{c|cccccccccccccccccccccccccccccccccc}	
		b_{j+\frac{1}{2}}  & \phantom{\bul} &{*} &  \phantom{\bul} &{*} &  \phantom{\bul} &{*}&  \phantom{\bul} &{*}&  \phantom{\bul} &{*}&  \phantom{\bul}&{*} &
		 \phantom{\bul} &{*}& \phantom{\bul} &{*}&  \phantom{\bul} &{*}&
		 \phantom{\bul} &{*}&  \phantom{\bul} &{*} &  \phantom{\bul} &{*} & \phantom{\bul}&{*}& \phantom{\bul} &{\movl} &  \phantom{\bul} & {*} \\
		\hline
		p_{j}  & \bul &\phantom{*} & \bul &\phantom{*} & \bul &\phantom{*}& \bul &\phantom{*}& \bul &\phantom{*}& \iga&\phantom{*} & \wga &\phantom{*}& \sga &\phantom{*}&  \iga &\phantom{*}&
		\iga &\phantom{*}& \sga &\phantom{*} & \wga &\phantom{*} & 
		\iga &\phantom{*}&\bul&\phantom{*}& \bul &\phantom{*}&
		\\		
		d_{j}  & 0  &\phantom{*}    & 1  &\phantom{*}     & 0   &\phantom{*}  & 0   &\phantom{*}  & 0   &\phantom{*} &1   &\phantom{*}  & 
		w_1  &\phantom{*} & w_2  &\phantom{*} & w_3  &\phantom{*} & 1   &\phantom{*}&
		0   &\phantom{*}  & 0   &\phantom{*}  & 0 &\phantom{*}   & 1   &\phantom{*}    	& 0   &\phantom{*}	
 	\end{array} 
\end{eqnarray*}
The solid left-triangle moves one step forward and turns into an empty left-triangle (via the rule 1a):
\begin{eqnarray*}
[30]\qquad	\begin{array}{c|cccccccccccccccccccccccccccccccccc}	
		b_{j+\frac{1}{2}}  & \phantom{\bul} &{*} &  \phantom{\bul} &{*} &  \phantom{\bul} &{*}&  \phantom{\bul} &{*}&  \phantom{\bul} &{*}&  \phantom{\bul}&{*} &
		 \phantom{\bul} &{*}& \phantom{\bul} &{*}&  \phantom{\bul} &{*}&
		 \phantom{\bul} &{*}&  \phantom{\bul} &{*} &  \phantom{\bul} &{*} & \phantom{\bul}&{\movle}& \phantom{\bul} &{*} &  \phantom{\bul} & {*} \\
		\hline
		p_{j}  & \bul &\phantom{*} & \bul &\phantom{*} & \bul &\phantom{*}& \bul &\phantom{*}& \bul &\phantom{*}& \iga&\phantom{*} & \wga &\phantom{*}& \sga &\phantom{*}&  \iga &\phantom{*}&
		\iga &\phantom{*}& \sga &\phantom{*} & \wga &\phantom{*} & 
		\iga &\phantom{*}&\bul&\phantom{*}& \bul &\phantom{*}&
		\\		
		d_{j}  & 0  &\phantom{*}    & 1  &\phantom{*}     & 0   &\phantom{*}  & 0   &\phantom{*}  & 0   &\phantom{*} &1   &\phantom{*}  & 
		w_1  &\phantom{*} & w_2  &\phantom{*} & w_3  &\phantom{*} & 1   &\phantom{*}&
		0   &\phantom{*}  & 0   &\phantom{*}  & 0 &\phantom{*}   & 1   &\phantom{*}    	& 0   &\phantom{*}	
 	\end{array} 
\end{eqnarray*}
The empty left-triangle can move to the left step by step via the rule 1c, until the configuration becomes:
\begin{eqnarray*}
[38]\qquad	\begin{array}{c|cccccccccccccccccccccccccccccccccc}	
		b_{j+\frac{1}{2}}  & \phantom{\bul} &{*} &  \phantom{\bul} &{*} &  \phantom{\bul} &{*}&  \phantom{\bul} &{*}&  \phantom{\bul} &{\movle}&  \phantom{\bul}&{*} &
		 \phantom{\bul} &{*}& \phantom{\bul} &{*}&  \phantom{\bul} &{*}&
		 \phantom{\bul} &{*}&  \phantom{\bul} &{*} &  \phantom{\bul} &{*} & \phantom{\bul}&{*}& \phantom{\bul} &{*} &  \phantom{\bul} & {*} \\
		\hline
		p_{j}  & \bul &\phantom{*} & \bul &\phantom{*} & \bul &\phantom{*}& \bul &\phantom{*}& \bul &\phantom{*}& \iga&\phantom{*} & \wga &\phantom{*}& \sga &\phantom{*}&  \iga &\phantom{*}&
		\iga &\phantom{*}& \sga &\phantom{*} & \wga &\phantom{*} & 
		\iga &\phantom{*}&\bul&\phantom{*}& \bul &\phantom{*}&
		\\		
		d_{j}  & 0  &\phantom{*}    & 1  &\phantom{*}     & 0   &\phantom{*}  & 0   &\phantom{*}  & 0   &\phantom{*} &1   &\phantom{*}  & 
		w_1  &\phantom{*} & w_2  &\phantom{*} & w_3  &\phantom{*} & 1   &\phantom{*}&
		0   &\phantom{*}  & 0   &\phantom{*}  & 0 &\phantom{*}   & 1   &\phantom{*}    	& 0   &\phantom{*}	
 	\end{array} 
\end{eqnarray*}
The empty left-triangle moves to the left and turns into a turn-around symbol (via the rule 1b):
\begin{eqnarray*}
[39]\qquad	\begin{array}{c|cccccccccccccccccccccccccccccccccc}	
		b_{j+\frac{1}{2}}  & \phantom{\bul} &{*} &  \phantom{\bul} &{*} &  \phantom{\bul} &{*}&  \phantom{\bul} &{\tur}&  \phantom{\bul} &{*}&  \phantom{\bul}&{*} &
		 \phantom{\bul} &{*}& \phantom{\bul} &{*}&  \phantom{\bul} &{*}&
		 \phantom{\bul} &{*}&  \phantom{\bul} &{*} &  \phantom{\bul} &{*} & \phantom{\bul}&{*}& \phantom{\bul} &{*} &  \phantom{\bul} & {*} \\
		\hline
		p_{j}  & \bul &\phantom{*} & \bul &\phantom{*} & \bul &\phantom{*}& \bul &\phantom{*}& \bul &\phantom{*}& \iga&\phantom{*} & \wga &\phantom{*}& \sga &\phantom{*}&  \iga &\phantom{*}&
		\iga &\phantom{*}& \sga &\phantom{*} & \wga &\phantom{*} & 
		\iga &\phantom{*}&\bul&\phantom{*}& \bul &\phantom{*}&
		\\		
		d_{j}  & 0  &\phantom{*}    & 1  &\phantom{*}     & 0   &\phantom{*}  & 0   &\phantom{*}  & 0   &\phantom{*} &1   &\phantom{*}  & 
		w_1  &\phantom{*} & w_2  &\phantom{*} & w_3  &\phantom{*} & 1   &\phantom{*}&
		0   &\phantom{*}  & 0   &\phantom{*}  & 0 &\phantom{*}   & 1   &\phantom{*}    	& 0   &\phantom{*}	
 	\end{array} 
\end{eqnarray*}
Because of two qubits nearby are $00$, the turn-around symbol becomes a right arrow (via the rule 3b):
\begin{eqnarray*}
[40]\qquad	\begin{array}{c|cccccccccccccccccccccccccccccccccc}	
		b_{j+\frac{1}{2}}  & \phantom{\bul} &{*} &  \phantom{\bul} &{*} &  \phantom{\bul} &{*}&  \phantom{\bul} &{\ff}&  \phantom{\bul} &{*}&  \phantom{\bul}&{*} &
		 \phantom{\bul} &{*}& \phantom{\bul} &{*}&  \phantom{\bul} &{*}&
		 \phantom{\bul} &{*}&  \phantom{\bul} &{*} &  \phantom{\bul} &{*} & \phantom{\bul}&{*}& \phantom{\bul} &{*} &  \phantom{\bul} & {*} \\
		\hline
		p_{j}  & \bul &\phantom{*} & \bul &\phantom{*} & \bul &\phantom{*}& \bul &\phantom{*}& \bul &\phantom{*}& \iga&\phantom{*} & \wga &\phantom{*}& \sga &\phantom{*}&  \iga &\phantom{*}&
		\iga &\phantom{*}& \sga &\phantom{*} & \wga &\phantom{*} & 
		\iga &\phantom{*}&\bul&\phantom{*}& \bul &\phantom{*}&
		\\		
		d_{j}  & 0  &\phantom{*}    & 1  &\phantom{*}     & 0   &\phantom{*}  & 0   &\phantom{*}  & 0   &\phantom{*} &1   &\phantom{*}  & 
		w_1  &\phantom{*} & w_2  &\phantom{*} & w_3  &\phantom{*} & 1   &\phantom{*}&
		0   &\phantom{*}  & 0   &\phantom{*}  & 0 &\phantom{*}   & 1   &\phantom{*}    	& 0   &\phantom{*}	
 	\end{array} 
\end{eqnarray*}
The right arrow moves one step to the right and becomes an empty right-triangle (via the rule 2b):
\begin{eqnarray*}
[41]\qquad	\begin{array}{c|cccccccccccccccccccccccccccccccccc}	
		b_{j+\frac{1}{2}}  & \phantom{\bul} &{*} &  \phantom{\bul} &{*} &  \phantom{\bul} &{*}&  \phantom{\bul} &{*}&  \phantom{\bul} &{\mov}&  \phantom{\bul}&{*} &
		 \phantom{\bul} &{*}& \phantom{\bul} &{*}&  \phantom{\bul} &{*}&
		 \phantom{\bul} &{*}&  \phantom{\bul} &{*} &  \phantom{\bul} &{*} & \phantom{\bul}&{*}& \phantom{\bul} &{*} &  \phantom{\bul} & {*} \\
		\hline
		p_{j}  & \bul &\phantom{*} & \bul &\phantom{*} & \bul &\phantom{*}& \bul &\phantom{*}& \bul &\phantom{*}& \iga&\phantom{*} & \wga &\phantom{*}& \sga &\phantom{*}&  \iga &\phantom{*}&
		\iga &\phantom{*}& \sga &\phantom{*} & \wga &\phantom{*} & 
		\iga &\phantom{*}&\bul&\phantom{*}& \bul &\phantom{*}&
		\\		
		d_{j}  & 0  &\phantom{*}    & 1  &\phantom{*}     & 0   &\phantom{*}  & 0   &\phantom{*}  & 0   &\phantom{*} &1   &\phantom{*}  & 
		w_1  &\phantom{*} & w_2  &\phantom{*} & w_3  &\phantom{*} & 1   &\phantom{*}&
		0   &\phantom{*}  & 0   &\phantom{*}  & 0 &\phantom{*}   & 1   &\phantom{*}    	& 0   &\phantom{*}	
 	\end{array} 
\end{eqnarray*}
Unlike the solid right-triangle, the empty right-triangle does not induce gate operation and simply moves one step forward and turns into a right arrow:
\begin{eqnarray*}
[42]\qquad	\begin{array}{c|cccccccccccccccccccccccccccccccccc}	
		b_{j+\frac{1}{2}}  & \phantom{\bul} &{*} &  \phantom{\bul} &{*} &  \phantom{\bul} &{*}&  \phantom{\bul} &{*}&  \phantom{\bul} &{\ff}&  \phantom{\bul}&{*} &
		 \phantom{\bul} &{*}& \phantom{\bul} &{*}&  \phantom{\bul} &{*}&
		 \phantom{\bul} &{*}&  \phantom{\bul} &{*} &  \phantom{\bul} &{*} & \phantom{\bul}&{*}& \phantom{\bul} &{*} &  \phantom{\bul} & {*} \\
		\hline
		p_{j}  & \bul &\phantom{*} & \bul &\phantom{*} & \bul &\phantom{*}& \bul &\phantom{*}&  \iga&\phantom{*}& \bul&\phantom{*} & \wga &\phantom{*}& \sga &\phantom{*}&  \iga &\phantom{*}&
		\iga &\phantom{*}& \sga &\phantom{*} & \wga &\phantom{*} & 
		\iga &\phantom{*}&\bul&\phantom{*}& \bul &\phantom{*}&
		\\		
		d_{j}  & 0  &\phantom{*}    & 1  &\phantom{*}     & 0   &\phantom{*}  & 0   &\phantom{*}  & 0   &\phantom{*} &1   &\phantom{*}  & 
		w_1  &\phantom{*} & w_2  &\phantom{*} & w_3  &\phantom{*} & 1   &\phantom{*}&
		0   &\phantom{*}  & 0   &\phantom{*}  & 0 &\phantom{*}   & 1   &\phantom{*}    	& 0   &\phantom{*}	
 	\end{array} 
\end{eqnarray*}
The previous two steps repeat a few times and we arrive at
\begin{eqnarray*}
	[55]\qquad\begin{array}{c|cccccccccccccccccccccccccccccccccc}	
		b_{j+\frac{1}{2}}  & \phantom{\bul} &{*} &  \phantom{\bul} &{*} &  \phantom{\bul} &{*}&  \phantom{\bul} &{*}&  \phantom{\bul} &{*}&  \phantom{\bul}&{*} &
		 \phantom{\bul} &{*}& \phantom{\bul} &{*}&  \phantom{\bul} &{*}&
		 \phantom{\bul} &{*}&  \phantom{\bul} &{*} &  \phantom{\bul} &{\mov} & \phantom{\bul}&{*}& \phantom{\bul} &{*} &  \phantom{\bul} & {*} \\
		\hline
		p_{j}  & \bul &\phantom{*} & \bul &\phantom{*} & \bul &\phantom{*}&   \bul&\phantom{*} & \iga&\phantom{*}& \wga &\phantom{*}& \sga &\phantom{*}&  \iga &\phantom{*}&
		\iga &\phantom{*}& \sga &\phantom{*} & \wga &\phantom{*} & \bul &\phantom{*}&
		\iga &\phantom{*}&\bul&\phantom{*}& \bul &\phantom{*}&
		\\		
		d_{j}  & 0  &\phantom{*}    & 1  &\phantom{*}     & 0   &\phantom{*}  & 0   &\phantom{*}  & 0   &\phantom{*} &1   &\phantom{*}  & 
		w_1  &\phantom{*} & w_2  &\phantom{*} & w_3  &\phantom{*} & 1   &\phantom{*}&
		0   &\phantom{*}  & 0   &\phantom{*}  & 0 &\phantom{*}   & 1   &\phantom{*}    	& 0   &\phantom{*}	
 	\end{array} 
\end{eqnarray*}
The empty right-triangle moves and becomes a right arrow:
\begin{eqnarray*}
[56]\qquad	\begin{array}{c|cccccccccccccccccccccccccccccccccc}	
		b_{j+\frac{1}{2}}  & \phantom{\bul} &{*} &  \phantom{\bul} &{*} &  \phantom{\bul} &{*}&  \phantom{\bul} &{*}&  \phantom{\bul} &{*}&  \phantom{\bul}&{*} &
		 \phantom{\bul} &{*}& \phantom{\bul} &{*}&  \phantom{\bul} &{*}&
		 \phantom{\bul} &{*}&  \phantom{\bul} &{*} &  \phantom{\bul} &{\ff} & \phantom{\bul}&{*}& \phantom{\bul} &{*} &  \phantom{\bul} & {*} \\
		\hline
		p_{j}  & \bul &\phantom{*} & \bul &\phantom{*} & \bul &\phantom{*}&   \bul&\phantom{*} & \iga&\phantom{*}& \wga &\phantom{*}& \sga &\phantom{*}&  \iga &\phantom{*}&
		\iga &\phantom{*}& \sga &\phantom{*} & \wga &\phantom{*} & 
		\iga &\phantom{*}&\bul &\phantom{*}&\bul&\phantom{*}& \bul &\phantom{*}&
		\\		
		d_{j}  & 0  &\phantom{*}    & 1  &\phantom{*}     & 0   &\phantom{*}  & 0   &\phantom{*}  & 0   &\phantom{*} &1   &\phantom{*}  & 
		w_1  &\phantom{*} & w_2  &\phantom{*} & w_3  &\phantom{*} & 1   &\phantom{*}&
		0   &\phantom{*}  & 0   &\phantom{*}  & 0 &\phantom{*}   & 1   &\phantom{*}    	& 0   &\phantom{*}	
 	\end{array} 
\end{eqnarray*}
The right arrow moves and becomes an empty right-triangle:
\begin{eqnarray*}
[57]\qquad	\begin{array}{c|cccccccccccccccccccccccccccccccccc}	
		b_{j+\frac{1}{2}}  & \phantom{\bul} &{*} &  \phantom{\bul} &{*} &  \phantom{\bul} &{*}&  \phantom{\bul} &{*}&  \phantom{\bul} &{*}&  \phantom{\bul}&{*} &
		 \phantom{\bul} &{*}& \phantom{\bul} &{*}&  \phantom{\bul} &{*}&
		 \phantom{\bul} &{*}&  \phantom{\bul} &{*} &  \phantom{\bul} &{*} & \phantom{\bul}&{\mov}& \phantom{\bul} &{*} &  \phantom{\bul} & {*} \\
		\hline
		p_{j}  & \bul &\phantom{*} & \bul &\phantom{*} & \bul &\phantom{*}&   \bul&\phantom{*} & \iga&\phantom{*}& \wga &\phantom{*}& \sga &\phantom{*}&  \iga &\phantom{*}&
		\iga &\phantom{*}& \sga &\phantom{*} & \wga &\phantom{*} & 
		\iga &\phantom{*}&\bul &\phantom{*}&\bul&\phantom{*}& \bul &\phantom{*}&
		\\		
		d_{j}  & 0  &\phantom{*}    & 1  &\phantom{*}     & 0   &\phantom{*}  & 0   &\phantom{*}  & 0   &\phantom{*} &1   &\phantom{*}  & 
		w_1  &\phantom{*} & w_2  &\phantom{*} & w_3  &\phantom{*} & 1   &\phantom{*}&
		0   &\phantom{*}  & 0   &\phantom{*}  & 0 &\phantom{*}   & 1   &\phantom{*}    	& 0   &\phantom{*}	
 	\end{array} 
\end{eqnarray*}
Via the rule 5b, the empty right-triangle turns into a solid left-triangle:
\begin{eqnarray*}
[58]\qquad	\begin{array}{c|cccccccccccccccccccccccccccccccccc}	
		b_{j+\frac{1}{2}}  & \phantom{\bul} &{*} &  \phantom{\bul} &{*} &  \phantom{\bul} &{*}&  \phantom{\bul} &{*}&  \phantom{\bul} &{*}&  \phantom{\bul}&{*} &
		 \phantom{\bul} &{*}& \phantom{\bul} &{*}&  \phantom{\bul} &{*}&
		 \phantom{\bul} &{*}&  \phantom{\bul} &{*} &  \phantom{\bul} &{*}& \phantom{\bul} &{\movl} & \phantom{\bul}&{*} &  \phantom{\bul} & {*} \\
		\hline
		p_{j}  & \bul &\phantom{*} & \bul &\phantom{*} & \bul &\phantom{*}&   \bul&\phantom{*} & \iga&\phantom{*}& \wga &\phantom{*}& \sga &\phantom{*}&  \iga &\phantom{*}&
		\iga &\phantom{*}& \sga &\phantom{*} & \wga &\phantom{*} & 
		\iga &\phantom{*}&\bul &\phantom{*}&\bul&\phantom{*}& \bul &\phantom{*}&
		\\		
		d_{j}  & 0  &\phantom{*}    & 1  &\phantom{*}     & 0   &\phantom{*}  & 0   &\phantom{*}  & 0   &\phantom{*} &1   &\phantom{*}  & 
		w_1  &\phantom{*} & w_2  &\phantom{*} & w_3  &\phantom{*} & 1   &\phantom{*}&
		0   &\phantom{*}  & 0   &\phantom{*}  & 0 &\phantom{*}   & 1   &\phantom{*}    	& 0   &\phantom{*}	
 	\end{array} 
\end{eqnarray*}
It then moves one step to the left and turns into an empty left-triangle:
\begin{eqnarray*}
[59]\qquad	\begin{array}{c|cccccccccccccccccccccccccccccccccc}	
		b_{j+\frac{1}{2}}  & \phantom{\bul} &{*} &  \phantom{\bul} &{*} &  \phantom{\bul} &{*}&  \phantom{\bul} &{*}&  \phantom{\bul} &{*}&  \phantom{\bul}&{*} &
		 \phantom{\bul} &{*}& \phantom{\bul} &{*}&  \phantom{\bul} &{*}&
		 \phantom{\bul} &{*}&  \phantom{\bul} &{*} &  \phantom{\bul} &{\movle}& \phantom{\bul} &{*} & \phantom{\bul}&{*} &  \phantom{\bul} & {*} \\
		\hline
		p_{j}  & \bul &\phantom{*} & \bul &\phantom{*} & \bul &\phantom{*}&   \bul&\phantom{*} & \iga&\phantom{*}& \wga &\phantom{*}& \sga &\phantom{*}&  \iga &\phantom{*}&
		\iga &\phantom{*}& \sga &\phantom{*} & \wga &\phantom{*} & 
		\iga &\phantom{*}&\bul &\phantom{*}&\bul&\phantom{*}& \bul &\phantom{*}&
		\\		
		d_{j}  & 0  &\phantom{*}    & 1  &\phantom{*}     & 0   &\phantom{*}  & 0   &\phantom{*}  & 0   &\phantom{*} &1   &\phantom{*}  & 
		w_1  &\phantom{*} & w_2  &\phantom{*} & w_3  &\phantom{*} & 1   &\phantom{*}&
		0   &\phantom{*}  & 0   &\phantom{*}  & 0 &\phantom{*}   & 1   &\phantom{*}    	& 0   &\phantom{*}	
 	\end{array} 
\end{eqnarray*}
After a few rounds, we finally arrive at the final state $|\psi_T\rangle$ via the transition rules and all the gates have been applied: 
\begin{eqnarray*}
[154]\qquad	\begin{array}{c|cccccccccccccccccccccccccccccccccc}	
		b_{j+\frac{1}{2}}  & \phantom{\bul} &{\movle} &  \phantom{\bul} &{*} &  \phantom{\bul} &{*}&  \phantom{\bul} &{*}&  \phantom{\bul} &{*}&  \phantom{\bul}&{*} &
		 \phantom{\bul} &{*}& \phantom{\bul} &{*}&  \phantom{\bul} &{*}&
		 \phantom{\bul} &{*}&  \phantom{\bul} &{*} &  \phantom{\bul} &{*}& \phantom{\bul} &{*} & \phantom{\bul}&{*} &  \phantom{\bul} & {*} \\
		\hline
		p_{j}  &   \bul&\phantom{*} & \iga&\phantom{*}& \wga &\phantom{*}& \sga &\phantom{*}&  \iga &\phantom{*}&
		\iga &\phantom{*}& \sga &\phantom{*} & \wga &\phantom{*} & 
		\iga &\phantom{*}&\bul &\phantom{*}&\bul&\phantom{*}& \bul &\phantom{*}&\bul &\phantom{*} & \bul &\phantom{*} & \bul &\phantom{*}& 
		\\		
		d_{j}  & 0  &\phantom{*}    & 1  &\phantom{*}     & 0   &\phantom{*}  & 0   &\phantom{*}  & 0   &\phantom{*} &1   &\phantom{*}  & 
		w_1  &\phantom{*} & w_2  &\phantom{*} & w_3  &\phantom{*} & 1   &\phantom{*}&
		0   &\phantom{*}  & 0   &\phantom{*}  & 0 &\phantom{*}   & 1   &\phantom{*}    	& 0   &\phantom{*}	
 	\end{array} 
\end{eqnarray*}
There is no futher forward transition. With the above examples and the transition rules, we can count the total number of transitions made in getting to the last configuration is $T=6+ (n+1)\Big(3R(R-1)(n+1)+9R-5\Big)$ and there are in total $T+1$ history states. The step number (counting from zero) is indicated in the square brackets $[\,]$ in the above configurations.
As remarked earlier, the quantum computation will be executed not by discrete transitions, but via the continuous time evolution via the Hamiltonian: $\exp(-iHt)$. The construction for the Hamiltonian is similar to that in Sec.~\ref{sec:5state}, we will not explicitly write down all terms. All that is needed is the effective Hamiltonian in the subspace of the history states. We will analyze the probability of arriving at a desired computation using the history-state basis in Sec.~\ref{sec:Prob}. 

\end{widetext}
\noindent {\bf Periodic boundary condition}. We have been using an open boundary condition, namely, the first site is not connected to the last site. What if our geometry is a circle? In this case, the transition will not terminate, but will continue forever, with gates being applied multiple times. To prevent this from happening, we simply add an additional column at the last site with $*$ replaced by a state $X$. There is no transition rule regarding $X$, so the transition will terminate once the symbol $\movle$ is one site next to $X$ from the other end. But this raises dimension $d$ on the half-integer site to 9. We note that for the periodic boundary condition the local dimension of the 2-local 10-state construction by Nagaj and Wocjan~\cite{NagajWocjan} will need to increase to 12  and their 20-state construction needs to be modified to have 22 states. 
\section{Probability analysis}
\label{sec:Prob}
One common feature of both the above constructions is that with the proper initial condition, there is only one forward transition rule that applies, except at the last configuration. At any stage, there is only one backward transition rule that applies, except at the initial configuration. We will refer to the quantum states associated with these configurations linked via transition rules as the history states.
In the basis of the valid history states $\ket{\psi_t}$ ($t=0\dots T$ via the transition rules), the effective Hamiltonian is
\begin{equation}
H_{\rm eff} =-\sum_{t=0}^{T-1} |\psi_{t+1}\rangle\langle \psi_{t}|+|\psi_{t}\rangle\langle \psi_{t+1}|.
\end{equation}
For simplicity, we will also denote $|\psi_{t}\rangle$ simply by $|t\rangle$, and the transition probability to arrive at the state $|m\rangle$ starting from $|0\rangle$ after evolving for time duration $\tau$ is
\begin{equation}
 p_{\tau}(m|0) = \left| 
    \bra{m} e^{-iH_{\rm eff} \tau} \ket{0}
   \right|^2.
  \end{equation}
  As shown by Nagaj and Wocjan~\cite{NagajWocjan}, for such a one-dimensional quantum walk Hamiltonian, the probability of arriving at states $m>T/q$ (where $q$ is an positive integer greater than unity) is 
  \begin{eqnarray*}
 P_{m> T/q}= \sum_{m> T/q} \frac{1}{\tau_0} \int_{0}^{\tau_0}d\tau\, p_{\tau}(m|0) \ge \frac{q-1}{q} 
   -{\cal O}(T/\tau_0).\end{eqnarray*}
   This shows that the averaged total probability of ending up at a history state with label $m> T/q$ is very high. As mentioned in the Introduction, the trick to boost the success probability of completing the desired computation is to pad sufficient identity gates so that for any $m> T/q$ the desired circuit has been executed. Nagaj and Wocjan simply took $q=6$. Note that $\tau_0$ only needs to be chosen so that $T/\tau_0$ is small, for example, $\tau_0={\cal O}(T\log T)$. 
   
\smallskip \noindent {\bf Measurement}.  As implied by the above averaged probability, the measurement scheme is to fix an appropriate $\tau_0$ and randomly at a time $\tau$ between $0$ and $\tau_0$, perform measurement on all sites in the computational basis and check the configuration to see if a desired history state $m>T/q$ is obtained. If so, the computation has passed the desired part and the remaining gates are all identity, and the output of the computational qubits is accepted. If not, the measurement outcome is thrown away, and the whole computation is re-started. But the analysis above shows that with  sufficient pad of identity gates and an appropriately chosen $\tau_0$, the probability of completing the quantum computation is high. 
   
\section{Concluding remarks}
\label{sec:con}
We address the compromise between the locality $k$ and the local Hilbert space dimension $d$ for one-dimensional Hamiltonian quantum computation. 
Specifically, we provide a construction of Hamiltonian quantum computer for $k=3$ with $d=5$.  One implication is that simulating the dynamics for 1D chains of spin-2 particles is BQP-complete as it would allow us to simulate a quantum computer. This construction, together with the previous ones, gives rise to a delineation of the border between easy and hard one-dimensional Hamiltonians in terms of the complexity class BQP. It is possible that further improvement of the boundary can be made.  Imposing translation invariance  increases the required local dimension $d$. We thus also construct another 3-local ($k=3$) Hamiltonian that is invariant under translation of two sites but that requires $d$ to be 8. Simulating the dynamics for such translationally invariant Hamiltonians is also a BQP-complete task.
Correspondingly, there is also an easy-hard boundary on the locality $k$ vs local dimension $d$ plane for 1D translationally invariant Hamiltonians. 

We do not know whether our constructions are optimal, namely whether the local Hilbert space dimension $d$ is as small as it can be while maintaining the universality. For the 5-state construction, we believe it is likely the case if one insists that the local dimension on every site be the same. On each site of one sub-lattice, there are two different kinds of qubits ($\gate$ and $\qubit$), and the gate application exploits the two kinds of qubits. The additional state $\blank$ (unborn/dead) is also necessary. If we try to use only one kind of the qubit (hence reducing $d_A$ to $3$), we have to increase the local dimension at the other sub-lattice to $d_B=7$ by adding two other symbols to enact the gate operations (the construction is not shown here). 
For the continuous-time quantum cellular automata, our 3-local 8-state construction can be regarded as an improvement from the 2-local 20-state construction by Nagaj and Wocjan. These two constructions both have unique forward and backward transitions and the effective Hamiltonians are the same as the 1D quantum walk. The 2-local 10-state construction of Nagaj and Wocjan does not have unique forward and backward transitions and its Hamiltonian does not correspond to a 1D quantum walk. But $d=10$  is the lowest known so far for translationally invariant 2-local Hamiltonians. It may be possible for $d=8$ in our 3-local case to be further reduced if one does not use Hamiltonians constructed from unique forward and backward transitions for the appropriate history states. But we have not found one that has a lower local dimension. Regarding the local dimensions of  3-local quantum cellular automata, we also have other constructions with mixed local dimensions. For example we have one construction with mixed dimensions $d_A=2$ and $d_B=14$, another construction with $d_A=5$ and $d_B=12$, and yet another one with $d_A=6$ and $d_B=10$. We do not list these other constructions here.

One can ask similarly the compromise between $k$ and $d$ for one-dimensional QMA-complete local Hamiltonian problems. The lowest local dimension for 2-local Hamiltonians is $d=8$ due to a work by Hallgren, Nagaj and  Narayanaswami~\cite{Hallgren}. This means that 6-local ($k=6$) qubit ($d=2$) Hamiltonian problems are already QMA-complete. For $k=4$, at most $d=3$ is needed for QMA-completeness. But for $k=3$, how much lower than 8 can $d$ be?  Our 3-local 5-state construction does not give rise to a Hamiltonian that is QMA-complete, as there are illegal configurations that remain zero energy even if we impose 3-local penalty terms.

\medskip \noindent {\bf Acknowledgment.}   This work was supported by the
National Science Foundation under Grant No. PHY 1314748. J.C.L. acknowledges support from the Simons Summer Research Program 2015 at the Stony Brook University, where part of the work was carried out.

\appendix
\section{Elementary proof of the universality of the $W$ gate}
\label{app:proof}
In this Appendix we will provide a proof that the $W$ gate itself is universal. In the process of building up to the proof, we will also review a proof by Aharaonov that the Hadamard and Toffoli gates constitute a universal set of gates~\cite{Aharonov}. The fact that the $W$ gate~(\ref{eqn:W}) alone is universal has been known, e.g., see Ref.~\cite{Shepherd}. We will not be concerned with the efficiency, and we will add the swap gate $S$ to our repertoire of gates, i.e., $\{S,W\}$. If one can apply $W$ gate between any pair of qubits (including the choice of which qubit being the control and which being the target), then the swap gate is not necessary, as it corresponds just to re-wiring of the circuit.  In our 1D continuous-time quantum cellular automaton, we fix the order of the control and target qubits, so we need the $S$ gate there. When necessary, we will use subscripts to denote the control $s$ and target $t$ in the $W$ gate: $W_{s\rightarrow t}$ or equivalently denote them inside the bracket: $W(s,t)$. 

Let us begin by making some simple observations. First, if we fixed the control qubit to be in $\ket{1}$ then the action of the $W$ gate on the target is the following Hadamard-like gate, denoted by $H_y$,
\begin{equation}
H_y=\frac{1}{\sqrt{2}}\left(
\begin{array}{cc}
1 & -1  \\
1 & 1 
\end{array}\right).
\end{equation}
Thus with an ancilla, we can use the $W$ gate to simulate the $H_y$ gate and include it in our repertoire. Furthermore, by a direct calculation, we have
\begin{equation}
W_{1\rightarrow2}^4 S\, W_{1\rightarrow2}^4 S\, W_{1\rightarrow2}^4=I_1\otimes Z_2,
\end{equation}
and thus we can generate a Pauli Z gate
\begin{equation}
Z=\left(
\begin{array}{cc}
1 & 0  \\
0 & -1 
\end{array}\right).
\end{equation}
The product of $H_y$ and $Z$ gives rise to the usual Hadamard gate $H$,
\begin{equation}
H\equiv H_y Z =\frac{1}{\sqrt{2}}\left(
\begin{array}{cc}
1 & 1  \\
1 & -1 
\end{array}\right).
\end{equation}
With the Hadamard gate and the $Z$ gate, we obtain the Pauli $X$ gate
\begin{equation}
X=H\,Z\,H=\left(
\begin{array}{cc}
0 & 1  \\
1 & 0 
\end{array}\right).
\end{equation}
One can also obtain the Control-NOT gate ($C_X$) via
\begin{equation}
C_X=W_{1\rightarrow2}^2S\,W_{1\rightarrow2}^6 S\, W_{1\rightarrow2}^2 S\, W_{1\rightarrow2}^6=\left(
\begin{array}{cccc}
1 & 0 & 0 & 0 \\
0 & 1 & 0 & 0 \\
0 & 0 & 0 & 1\\
0 & 0 &1 & 0
\end{array}\right),
\end{equation}
which can also allow us to get the $X$ gate. From the Hadamard gate $H$ and the $C_X$ gate, we can obtain the Control-Z gate $C_Z$. With the $X$ and $Z$ gates, we can obtain the $Y$ gate (the Pauli Y gate up to a factor of $i$) and its inverse: 
\begin{equation}
Y\equiv XZ=\left(\begin{array}{cc}
0 & -1  \\
1 & 0 
\end{array}\right), \quad Y^{-1} =\left(\begin{array}{cc}
0 & 1  \\
-1 & 0 
\end{array}\right).
\end{equation}

\begin{figure}
 \includegraphics[width=0.3\textwidth]{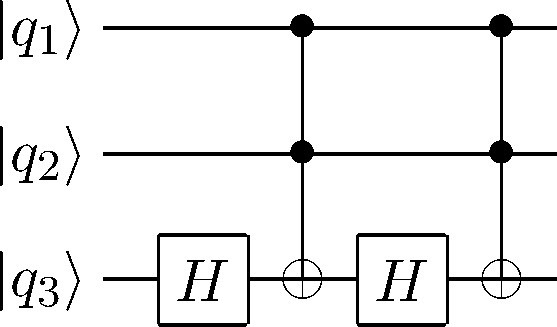}
        \caption{ The circuit to simulate the $\Lambda^2(Y)$ gate using the Toffoli and Hadamard gates. This construction relies on an identity that $Y=XHXH$.
         \label{fig:LY0} }
\end{figure}

\begin{figure}[t!]
 \includegraphics[width=0.45\textwidth]{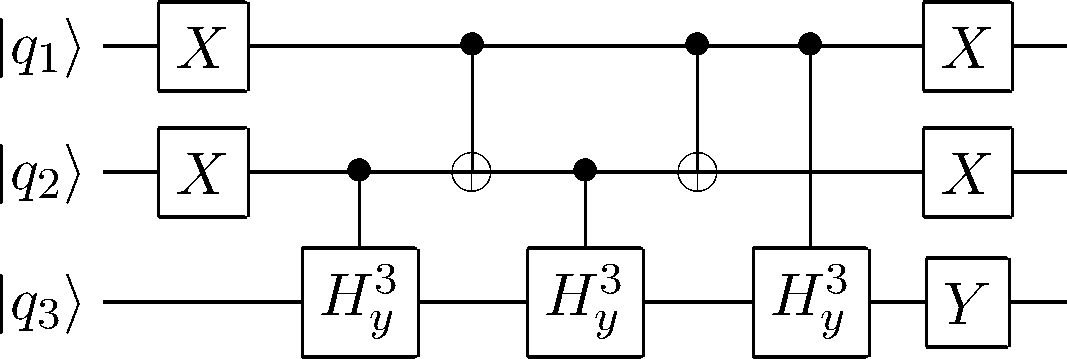}
        \caption{The circuit to simulate the $\Lambda^2(Y)$ gate via gates generated by $W$ gate or equivalently $C_{H_y}$ gate. Note that we have simplied the notation by replacing $W^3$ by a $C_{H_y^3}$ gate and that Control-NOT, $X$ and $Y$ gates can all be generated by the $W$ gate, as explained in the text.
         \label{fig:LY} }
\end{figure}

With the above observation, we are almost ready to prove the universality of the $W$ gate. But before we do that it is instructive to review the proof by Aharonov that the Toffoli gate $T$ and the Hadamard gate $H$ constitute a universal set of quantum gates~\cite{Aharonov}. In the proof she used a result by Kitaev~\cite{Kitaev97} that the Hadamard gate and the Control-Phase gate $\Lambda(P(i))$,
\begin{equation}
\Lambda(P(i))=\left(
\begin{array}{cccc}
1 & 0 & 0 & 0 \\
0 & 1 & 0 & 0 \\
0 & 0 & 1 & 0\\
0 & 0 &0 & i
\end{array}\right),
\end{equation}
 are also a universal set of gates.  She then applied the notion of encoded universality to turn the (two-qubit) Control-Phase gate into an equivalent three-qubit gate (with only real numbers) which is in fact the Control-Control-$Y$ gate, which we write explicitly below,
 \begin{equation}
\Lambda^2(Y)\equiv\left(
\begin{array}{cccccccc}
1 & 0 & 0 & 0 & 0 & 0 & 0& 0 \\
0 & 1 & 0 & 0 & 0 & 0 & 0& 0 \\
0 & 0 & 1 & 0 & 0 & 0 & 0& 0\\
0 & 0 &0 & 1 & 0 & 0 & 0& 0\\
0 & 0 &0 & 0 & 1 & 0 & 0& 0\\
0 & 0 &0 & 0 & 0 & 1 & 0& 0\\
0 & 0 &0 & 0 & 0 & 0 & 0& -1\\
0 & 0 &0 & 0 & 0 & 0 & 1 & 0
\end{array}\right).
\end{equation}
As shown in Fig.~\ref{fig:LY0}, this gate can be simulated by the Toffoli and the Hadamard gates, as
\begin{equation}
\Lambda^2(Y=XZ)=T(1,2,3)H(3)T(1,2,3)H(3),
\end{equation} 
where the numbers inside the brackets indicate which qubits are acted by the gate, and the explicit expression of the Toffoli gate (a.k.a. Control-Control-NOT gate) is
\begin{equation}
T\equiv\left(
\begin{array}{cccccccc}
1 & 0 & 0 & 0 & 0 & 0 & 0& 0 \\
0 & 1 & 0 & 0 & 0 & 0 & 0& 0 \\
0 & 0 & 1 & 0 & 0 & 0 & 0& 0\\
0 & 0 &0 & 1 & 0 & 0 & 0& 0\\
0 & 0 &0 & 0 & 1 & 0 & 0& 0\\
0 & 0 &0 & 0 & 0 & 1 & 0& 0\\
0 & 0 &0 & 0 & 0 & 0 & 0& 1\\
0 & 0 &0 & 0 & 0 & 0 & 1 & 0
\end{array}\right).
\end{equation}
Therefore, the Hadamard and Toffoli gates are universal.

Finally, we now return to finish the proof that the $W$ gate is itself universal. Since we already have the Hadamard gate in our repertoire, we prove the universality of the $W$ gate by showing that the gate $\Lambda^2(Y)$ can be simulated by the gates in the repertoire set generated from $W$, i.e.,
\begin{eqnarray}
\Lambda^2(Y)&=&X(1)X(2) Y(3) W(1,3)^3\,C_X(1,2)W(2,3)^3\nonumber\\
& &\,C_X(1,2)W(2,3)^3 \,X(2)X(1),
\end{eqnarray} 
the circuit for which is shown in Fig.~\ref{fig:LY}.
This equality can be verified directly by matrix multiplication or by using the elementary gate multiplication and identities (see chapter 4 of Ref.~\cite{NielsenChuang}). More specifically, by checking all possible control bits ($00$, $01$, $10$, $11$) one can verify that the combination of the five control gates in the middle of the circuit (Fig.~\ref{fig:LY}) gives rise to a Control-Control-$Y^{-1}$ gate, but conditioned on the first two qubits not both being $0$. Including the $Y$ gate makes this block of gates become a Control-Control-$Y$ gate conditioned on the first two qubits being both $0$. The $X$ gates on both qubits before and after the previous block make the whole circuit become a Control-Control-$Y$ gate (conditioned on first two qubits being both 1), i.e., $\Lambda^2(Y)$, hence completing the proof.

\end{document}